\documentclass[journal]{IEEEtran}
\ifCLASSINFOpdf
  \usepackage[pdftex]{graphicx}
   \usepackage{epstopdf}
\else
\fi

\usepackage{mathrsfs}
\usepackage{pifont}
\usepackage{bbding}
\usepackage{amsmath}
\usepackage{tipa}
\usepackage{algorithm}
\usepackage{algorithmic}
\usepackage{mathrsfs}
\usepackage{pifont}
\usepackage{bbding}
\usepackage{amsmath}
\usepackage{color}
\usepackage{tipa}
\usepackage{algorithm}
\usepackage{algorithmic}
\usepackage{epsfig}
\usepackage{amssymb}
\usepackage{amsfonts}
\usepackage{epic}
\usepackage{graphicx}
\usepackage{curves}
\usepackage{subfigure}
\usepackage[justification=centering]{caption}

\ifCLASSINFOpdf
\else
\fi
\begin{document}
\newtheorem{theorem}{Theorem}
\newtheorem{proposition}{Proposition}
\newtheorem{definition}{Definition}
\newtheorem{lemma}{Lemma}
\newtheorem{corollary}{Corollary}
\newtheorem{remark}{Remark}
\newtheorem{construction}{Construction}

\newcommand{\supp}{\mathop{\rm supp}}
\newcommand{\sinc}{\mathop{\rm sinc}}
\newcommand{\spann}{\mathop{\rm span}}
\newcommand{\essinf}{\mathop{\rm ess\,inf}}
\newcommand{\esssup}{\mathop{\rm ess\,sup}}
\newcommand{\Lip}{\rm Lip}
\newcommand{\sign}{\mathop{\rm sign}}
\newcommand{\osc}{\mathop{\rm osc}}
\newcommand{\R}{{\mathbb{R}}}
\newcommand{\Z}{{\mathbb{Z}}}
\newcommand{\C}{{\mathbb{C}}}
\captionsetup{font={scriptsize}}

%=======================================================title==============================================================================
%\title{ Tone Reservation for PAPR Reduction Method Based on FISTA }
\title{Efficient Joint Precoding Design for  Wideband Intelligent Reflecting Surface-Assisted Cell-Free Network}
%===================================== ==================author information=================================================================
\author{{Yajun~Wang, Jinghan~Jiang, Xin Du, Zhuxian~Lian, Qingqing~Wu,~\IEEEmembership{Senior Member,~IEEE}, and Wen Chen,~\IEEEmembership{Senior Member,~IEEE}}% and Qinghua Liu}
%\author{{Author 1, Author 2,}
%and Chintha Tellambura,~\IEEEmembership{Fellow,~IEEE}}
        % <-this % stops a space
%\thanks{Yajun~Wang  is  with Department of Information and Computational Sciences.
%JiangSu  University of Science and Technology, Zhangjiagang, 215600, e-mail:
%wangyj1859@just.edu.cn.}
%\thanks{Xin~Du, Yajun~Wang, Zhuxian~Lian, Yinjie~Su, and Zhibin~Xie are with the Department of Electronic Engineering.
%Jiangsu  University of Science and Technology, Zhenjiang, 212003,
%e-mail: 199030008@stu.just.edu.cn, wangyj1859@just.edu.cn, zhuxianlian@just.edu.cn, yinjiesu@just.edu.cn, xiezhibin@just.edu.cn.}}
%\thanks{Qinghua~Liu  is  with Department of  Computer Science.
%JiangSu  University of Science and Technology, Zhenjiang, 212003, e-mail:
%liuqh@just.edu.cn.}
% <-this % stops a space
%\thanks{This work is supported in part by the National Natural Science Foundation of China under Grant 62001194, Grant 61371114, and Grant 61401180.}}%, and in part by the Natural Science Foundation of Jiangsu province of China under Project BK20190956.}}
\thanks{Yajun~Wang, Jinghan~Jiang, Xin Du and Zhuxian~Lian are with the Department of Electronic Engineering,
Jiangsu  University of Science and Technology, Zhenjiang, 212003, China (e-mail: wangyj1859@just.edu.cn; 231112201113@stu.just.edu.cn; 211210301406@stu.just.edu.cn; zhuxianlian@just.edu.cn).}
\thanks{Qingqing Wu and  Wen Chen are with the Department of Electronic Engineering, Shanghai Jiao Tong University, Shanghai, 200240, China,
 (e-mail: qingqingwu@sjtu.edu.cn; wenchen@sjtu.edu.cn).}
%\thanks{Qinghua~Liu  is  with Department of  Computer Science.
%JiangSu  University of Science and Technology, Zhenjiang, 212003, e-mail:
%liuqh@just.edu.cn.}
% <-this % stops a space
\thanks{This work was supported in part by FDCT
under Grant 0119/2020/A3; in part by GDST under Grant 2020B1212030003;
in part by STDF, Macau, under Grant 0036/2019/A1; in part by the National
Key Project 2020YFB1807700 and Project 2018YFB1801102; and in part by
NSFC under Grant 62001194, 61872184.}}
%by SEU SKL project
%\#W200907, by Huawei Funding \#YJCB2009024WL and \#YJCB2008048WL,
%and by National 973 project \#2009CB824900.}}

%Telephone: (800) 555--1212, Fax: (888) 555--1212}

%\markboth{IEEE Signal Processing Letters (Submitted) }{Shell
%\MakeLowercase{\textit{et al.}}: Bare Demo of IEEEtran.cls for
%Journals }
 %make the title area
\maketitle
%=======================================================abstract=================================================================
\begin{abstract}
In this paper, we propose an efficient joint precoding design method to maximize the weighted sum-rate in wideband intelligent reflecting surface (IRS)-assisted cell-free networks by jointly optimizing the active beamforming of base stations and the passive beamforming of IRS. Due to employing wideband transmissions, the frequency selectivity of IRSs has to been taken into account, whose response usually follows a Lorentzian-like profile. To address the high-dimensional non-convex optimization problem,
we employ a fractional programming approach to decouple the non-convex problem into subproblems for alternating optimization between active and passive beamforming. The active beamforming subproblem is addressed using the consensus alternating direction method of multipliers (CADMM) algorithm, while the passive beamforming subproblem is tackled using the accelerated projection gradient (APG) method and Flecher-Reeves conjugate gradient method (FRCG).  Simulation results demonstrate that our proposed approach achieves significant improvements in weighted sum-rate under various performance metrics compared to primal-dual subgradient (PDS) with ideal reflection matrix. This study provides valuable insights for computational complexity reduction  and network capacity enhancement.%power consumption minimization,
\end{abstract}

% IEEEtran.cls defaults to using math in the Abstract.


% \fi
\IEEEpeerreviewmaketitle
%=======================================================section1 introduction=======================================================
\section{Introduction}\label{sec:1}

Ultra-dense networking  technology (UDN) enhances network capacity by increasing the number of base stations (BSs) and utilizing small cells~\cite{IEEEconf:1}. In traditional cellular networks, a BS typically serves all users within a cell, leading to significant interference for users at the cell boundaries from neighboring cells~\cite{IEEEconf:2}. UDN addresses this issue by dividing the cell into smaller units through an increased number of BSs, thereby reducing the number of users per cell and improving network capacity and coverage. However, this approach also introduces challenges in terms of stronger inter-cell interference due to a higher BS density. Consequently, the potential improvement in network capacity is limited as interference between adjacent cells increases.

The design concept of cell-free networks~\cite{IEEEconf:3},~\cite{IEEEconf:4} diverges from that of traditional cellular networks. Cell-free networks prioritize user-centricity, enabling seamless collaboration among all BSs without the constraints of cell boundaries. This enhanced flexibility in resource allocation optimizes signal transmission and fosters efficient BS cooperation,  providing services to users in unrestricted areas. Consequently, this network design not only effectively mitigates inter-cell interference but also enhances network capacity. As a result, extensive research efforts have been dedicated to various aspects such as resource allocation~\cite{IEEEconf:5}, precoding/beamforming~\cite{IEEEconf:6}, and channel estimation~\cite{IEEEconf:7}.

In the current cell-free network~\cite{IEEEconf:8},~\cite{IEEEconf:9}, increasing network capacity typically requires the addition of more distributed or small BSs, which presents challenges in terms of high power consumption and cost. However, as an emerging technology, intelligent reflective surfaces (IRSs) offer a viable alternative solution to address this problem~\cite{IEEEconf:10},~\cite{IEEEconf:11}. In recent years, IRS technology has garnered significant attention and has become a prominent research area. IRSs are usually metasurfaces, comprising numerous low-cost passive components. IRS can manipulate signal reflections by adjusting the phase and amplitude of each reflecting element~\cite{IEEEconf:10}-\cite{IEEEconf:15}. This inherent flexibility enables IRS to alter transmission paths for signal focusing and orientation. Compared to the deployment of additional BSs, the integration of low-power IRS into cell-free networks can effectively modify the communication environment while minimizing the generation of heating noise~\cite{IEEEconf:16}-\cite{IEEEconf:20}. Therefore, leveraging IRS as a potential solution allows for increased network capacity without relying on extensive deployment of distributed BSs.

IRSs have been extensively studied in various domains, including one-bit symbol-level precoding [21], energy efficiency [22], channel estimation [23], [24], and joint beamforming [25]-[29]. Some of these investigations focus on the use of  IRS to enhance network capacity, particularly through the design of joint precoding. The researchers proposed an iterative algorithm that combines alternating optimization and penalty-based techniques to jointly optimize transmit beamforming at the BS and reflective beamforming at the IRS [30]. The objective is to minimize the total transmit power at the BS. In addition, passive beamforming and information transfer techniques in IRS-based multi-user multiple-input multiple-output (MU-MIMO) systems  were discussed in [31]. Furthermore, achievable rate maximization problems in spectrum-sharing MIMO systems have been addressed using a penalized dyadic decomposition-based gradient projection algorithm [32]. Moreover, researchers also addressed optimizations of energy efficiency and spectral efficiency  of wireless communication aided by IRS as well as UAV networks [33]-[35]. Maximization of the minimum signal-to-interference-plus-noise ratio (SINR) in IRS-assisted multiple-input single-output (MISO) systems has been investigated to ensure fairness among users [36], [37]. Joint precoding problems involving  multi-antenna, multi-BS, multi-IRS, multi-user, and multi-carrier scenarios were addressed in [38]. Furthermore, research works have explored various technologies combined with IRS, such as maximizing the signal-to-noise ratio (SNR) and designing hybrid beamforming for millimeter-wave (mmWave) communication systems assisted by IRS [39], [40], maximizing the sum-rate and power allocation in orthogonal frequency division multiplexing (OFDM) systems assisted by IRS [41],  minimizing the total transmit power for radar communication assisted by IRS [42], and maximizing energy efficiency for hybrid active passive IRS-assisted energy efficient wireless communication [43]. Cooperative passive beamforming design was also investigated in double-IRS assisted multi-user MIMO systems [44], [45]. The channel model and performance prediction of double-IRS assisted communications were also analyzed in [46]. The design of beamforming for IRS-assisted cell-free massive MIMO communication systems was also addressed [47]-[50].

Nevertheless, the aforementioned studies on IRS-assisted communication systems have concentrated on narrowband systems, wherein the IRS is modeled as an adjustable phase shifter. In the context of wideband transmission, it is necessary to consider the frequency selective response of metamaterials. A frequency-dependent model for the amplitude and phase shift variations of the IRS elements has recently been proposed, and a weighted sum-rate maximization problem with OFDM modulations has been addressed [51], [52].  Nevertheless, the frequency response of a metamaterial typically exhibits a Lorentzian form [53]. While the application of metasurfaces in wideband communication systems has been recently explored  as active massive MIMO antennas  [54]-[56], there has been a paucity of research on passive metamaterial-based IRSs [57].

In this study, wideband  IRS-assisted cell-free networks are investigated, wherein the frequency selective response of the IRS behaves a Lorentzian form. Our primary objective is to maximize the network capacity in wide-band scenarios for IRS-assisted cell-free networks. To achieve this, we address the problem of joint precoding design with the aim of maximizing the weighted sum-rate (WSR) across all users by optimizing active beamforming at the BSs and passive beamforming at the IRSs. While previous works [58]-[63], such as alternating optimization method (AO), prox-linear block coordinate descent (BCD), successive convex approximation (SCA) [58], alternating direction method of multipliers (ADMM) and iterative reflection coefficient updating (ICU) [59], also consider WSR maximization, they are specific cases compared to our study. These previous methods primarily focus on scenarios involving a single BS and a single IRS, as well as narrowband communications.

The previous study [38] also investigated the maximization of the WSR in narrowband communications and proposed a primal-dual subgradient  method (PDS) to address this problem. However, the PDS method involves iterative updates and lacks closed-form solutions for active precoding at BSs and passive precoding at IRSs, resulting in high computational complexity. Moreover, the PDS method only converges to a suboptimal solution, leaving room for further performance improvement. Motivated by these observations, we reexamine the WSR maximization problem in IRS-assisted cell-free networks with the objective of developing a low complexity algorithm that overcomes the limitations of PDS methods. This paper highlights the following main contributions.

$\bullet$ To address the formidable task of maximizing the WSR in non-convex optimization problems with high-dimensional variables, we employ a two-step approach. First, we leverage the Lagrangian dual transformation (LDT) and multidimensional complex quadratic transformation (MCQT) techniques proposed in [64], [65]. This technique effectively decouples the active precoding/beamforming at the BS and the passive precoding at the IRS, thereby simplifying the overall optimization problem. Subsequently, we utilize  the consensus alternating direction method of multipliers (CADMM) algorithm [66], and the accelerated projection gradient (APG) method [67], [68], as well as Flecher-Reeves conjugate gradient (FRCG) method  to solve  subproblems related to active beamforming and passive beamforming, respectively. By integrating Lagrange duality transformation, MCQT, CADMM algorithms, and APG-FRCG algorithms together, we can effectively handle both non-convexity and high-dimensionality inherent in WSR maximization problems. Finally, our combined algorithm is denoted as CADMM-APG-FRCG algorithm.

$\bullet$ A commonly employed approach to tackle the decoupling precoding problem involves transforming it into a quadratic constrained quadratic programming (QCQP) subproblem. However, solving each QCQP subproblem poses significant challenges. In contrast to the method proposed in [38], our CADMM-APG-FRCG algorithm addresses QCQP subproblems with multiple constraints by decomposing them into multiple QCQP problems, each having only one constraint (QCQP-1). This decomposition offers the advantage of optimally solving all QCQP-1 subproblems and obtaining a simple closed-form solution. Unlike conventional methods, our algorithm eliminates the need for exhaustive search of optimal Lagrange dual variables, thereby avoiding convex or non-convex optimization problems during iteration. This algorithm  is  easy and efficient to implement, ensuring the effectiveness and excellent performance throughout the optimization process.

$\bullet$ The complexity analysis reveals that the proposed CADMM-APG-FRCG algorithm exhibits lower computational complexity compared to the existing PDS algorithm. Furthermore, the simulation results demonstrate a significant improvement in the WSR performance in all indices when  the CADMM-APG-FRCG algorithm is used instead of the PDS method. These findings indicate that the CADMM-APG-FRCG algorithm effectively optimizes the objective function and yields higher WSR values.

The remaining sections of this paper are organized as follows. Section II discusses the system model of IRS-assisted cell-free networks and presents the corresponding expression for the WSR maximization problem. In Section III, an alternative optimization framework is presented to address this problem. Section IV introduces the CADMM-APG-FRCG algorithm used for active and passive precoding. Simulations in Section V validate the performance of the presented CADMM-APG-FRCG algorithm for IRS-assisted cell-free networks. Finally, a conclusion is drawn in Section VI.

Notations: $\mathbb{R}$ and $\mathbb{C}$ denote  sets of real and complex numbers, respectively; $\left [ \cdot \right ]^{H}$, $\left [ \cdot \right ]^{T}$, $\left[\cdot\right]^{-1}$, and $\left [ \cdot  \right ]^{*}$  represent the conjugate-transpose, transpose, inverse, and conjugate  operations, respectively; $\left [ \mathbf{c}  \right ]_{j}$  is the $j$-th element of the vector $\mathbf{c}$; $\left [ \mathbf{B}  \right ]_{a,b}$  is the element in the $b$-th column and the $a$-th row of the matrix $\mathbf{B}$;  $E\left [ \cdot  \right ]$ is the expectation operator; $||\mathbf{c}||$  denotes the Euclidean norm of its argument; $diag\left ( \cdot  \right )$  denotes the diagonal operation; $\Re\{\cdot\}$  denotes the real part of its argument; $\otimes $ denotes the Kronecker product;  $\mathbf{0}_M$ stands for an $M \times M$ zero matrix. $\mathbf{I}_M$  denotes an $M\times M$ identity matrix.  $\mathbf{e}_m$  denotes a  primary vector with 1 at the $m$-th location, and $\mathbf{1}_M$ is an $M$-length vector with  elements being $1$. A circularly symmetric complex Gaussian ~random vector $\mathbf{c}$ with mean $\mathbf{0}$  and covariance matrix $\mathbf{\Pi}$ is denoted by $\mathbf{c}\sim\mathcal{CN}(\mathbf{0},\mathbf{\Pi})$. %$Tr\{\cdot\}$  denotes the trace of its argument; $ln\{\cdot\}$  denotes the natural logarithm; The sign $\bigotimes$ denotes  the Kronecker product of two matrices.

%=======================================================section2=====================================================================
\section{SYSTEM MODEL AND PROBLEM FORMULATION }\label{sec:2}

\subsection{System Model}
The main focus of this paper is to investigate a wideband cell-free network system that utilizes distributed BSs and IRSs to collectively serve a group of users, as shown in Fig.~1. The system consists of $N_b$ BSs, $N_c$ IRSs, $M$ subcarriers, and $K$ users with multiple antennas. To facilitate the analysis, we make certain assumptions. Each BS is equipped with $N_t$ antennas, each user has $N_r$ antennas, and each IRS is comprised of $R$ reflection units. We represent the index sets for the BSs, users, IRSs, subcarriers, and IRS elements as $\mathcal{N}_\mathcal{B}=\{1,\cdots,N_b\}$, $\mathcal{K}=\{1,\cdots,K\}$, $\mathcal{N}_{\mathcal{C}}=\{1,\cdots,N_c\}$, $\mathcal{M}=\{1,\cdots,M\}$, and $\mathcal{R}=\{1,\cdots,R\}$, respectively.

\begin{figure}
	\centering
	\includegraphics[width=3.5in,angle=0]{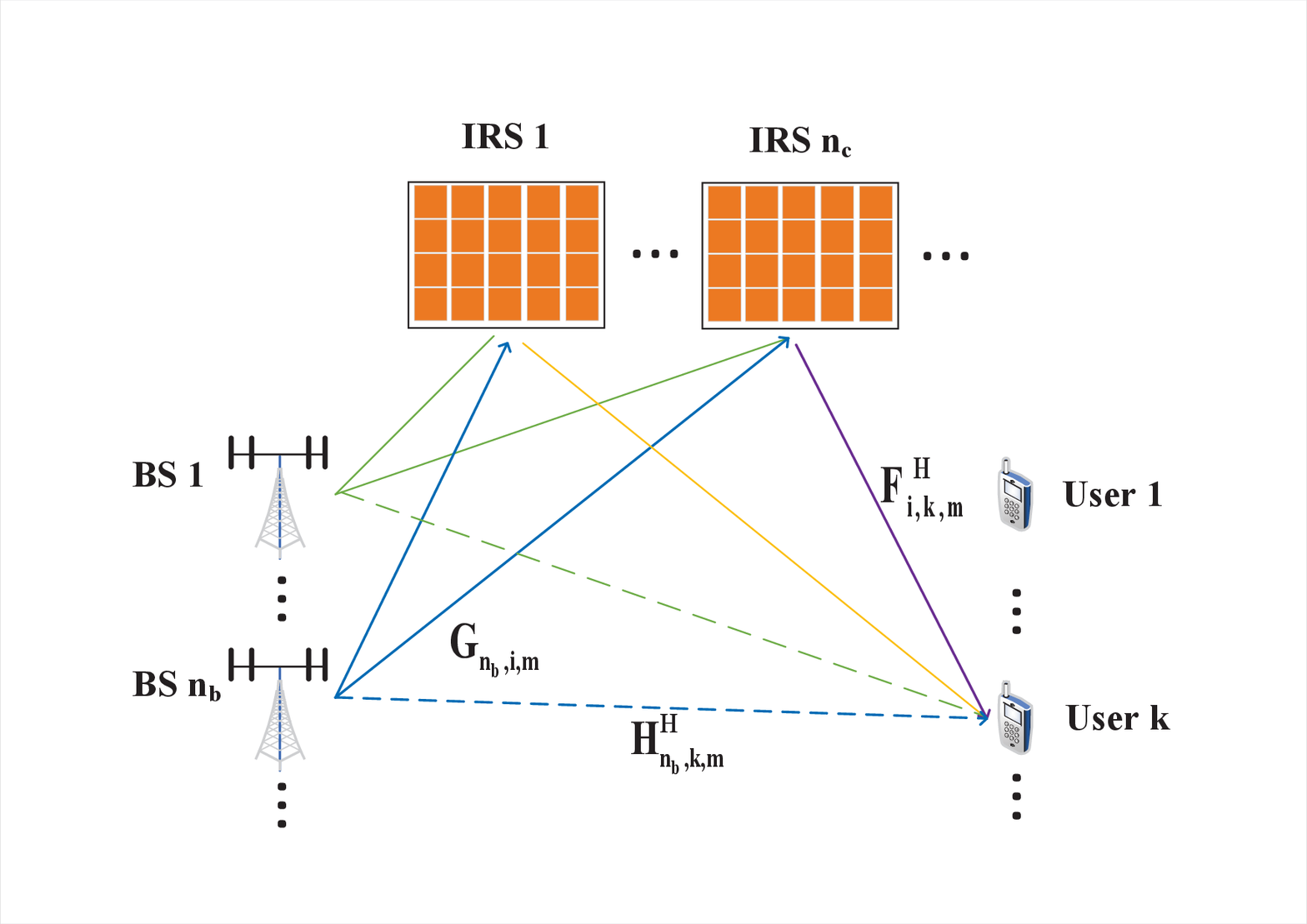}
	\caption{The downlink channels in the IRS-aided cell-free network.}
	\label{fig1}
\end{figure}
\subsubsection{Transmit Waveform Design}
\

Let $\mathbf{s}_{n_b, m}=[{s}_{n_b,m,1},\cdots,{s}_{n_b,m,K}]^{T}\in\mathbb{C}^{K}$, where ${s}_{n_b,m,k}$  represents the frequency domain symbol sent to the $k$-th user on the $m$-th subcarrier/tone from the $n_b$-th BS,  $\mathbf{s}_{n_b,m}\sim\mathcal{CN}(\mathbf{0}_{K},\mathbf{I}_K), \forall n_b\in\mathcal{N}_\mathcal{B}, \forall m\in\mathcal{M}$.
% We assume that the sent symbols have normalized power, i.e., $E\left [\mathbf{s}_{d}\mathbf{s}_{d}^{H}  \right ]=\mathbf{I}_K$, $\forall d\in\mathcal{D}$.
The precoded signal $\mathbf{x}_{n_b,m}$  at the $n_b$-th BS on the $m$-th subcarrier is given by
\begin{equation}\label{eq1}
\mathbf{x}_{n_b,m}=\sum_{j=1}^{K}\mathbf{w}_{n_b,m,j}s_{n_b,m,j},
\end{equation}
where $\mathbf{w}_{n_b,m,j}\in\mathbb{C}^{N_t}$ is the precoding vector at the $n_b$-th BS in the downlink.

Firstly, the precoded signal $\{\mathbf{x}_{n_b,m}\}_{m=1}^{M}$  on all $M$ tones at the $n_b$-th BS is transformed into the time domain using inverse discrete Fourier transform (IDFT). Subsequently, a cyclic prefix (CP) is added to the time-domain signals before converting them from the $N_t$ RF chain of the $n_b$-th BS to the radio frequency (RF) domain.

%First, we convert the precoded signal $\{\mathbf{x}_{n_b,z}\}_{z=1}^{Z}$  on all $Z$ tones at the $n_b$-th BS  into the time-domain by inverse discrete Fourier transform (IDFT). Then, the time-domain signals are added a cyclic prefix (CP) and converted from the $N_t$ RF chain of the $n_b$-th BS to the radio frequency (RF) domain.
\subsubsection{Channel Model}
\

In the presented IRS-assisted cell-free network, the channel between each BS and each user  is composed of three parts: a BS-user link, $N_c$ BS-IRS link, and $N_c$ IRS-user link. Consequently, the equivalent channel $\mathbf{h}_{n_b,k,m}^{H}\in\mathbb{C}^{N_r\times N_t}$  between the $n_b$-th BS and the $k$-th user on the $m$-th tone can be expressed as follows.
\begin{equation}\label{eq2}
\mathbf{\hat{H}}_{n_b,k,m}^{H}=\mathbf{H}_{n_b,k,m}^{H}+\sum_{i=1}^{N_c}\mathbf{F}_{i,k,m}^{H}\mathbf{\Phi}_{i,m}^H\mathbf{G}_{n_b,i,m},
\end{equation}
where $\mathbf{H}_{n_b,k,m}^{H}\in\mathbb{C}^{N_r\times N_t}$  , $\mathbf{G}_{n_b,i,m}\in\mathbb{C}^{R\times N_t}$ , $\mathbf{F}_{i,k,m}^{H}\in\mathbb{C}^{N_r\times R}$ represent the frequency domain channels on the $m$-th tone  from the $n_b$-th BS  to the $k$-th user, from the $n_b$-th BS  to the $i$-th IRS, and from the $i$-th IRS to the $k$-th user, respectively. $\mathbf{\Phi}_{i,m}\in\mathbb{C}^{R\times R}$  is the reflection matrix at the $i$-th IRS on the $m$-th tone, which is given by
\begin{equation}\label{eq3}
\mathbf{\Phi}_{i,m}=diag(\mathbf{\phi}_{i,1,m},\cdots,\mathbf{\phi}_{i,R,m}),\forall i\in\mathcal{N_C},
\end{equation}
where $\mathbf{\phi}_{i,r,m}$  is the reflection coefficient (RC) of the $r$-th element at the $i$-th IRS for the $m$-th tone.  Different from the  ideal model~\cite{IEEEconf:38}, we employ the metamaterial-based IRS model discussed in~\cite{IEEEconf:53}, i.e.,   $\mathbf{\phi}_{i,r,m}$ is modeled as a polarized dipole, and follow the following Lorentzian form.
\begin{equation}\label{eq4}
\begin{aligned}
\mathbf{\phi}_{i,r,m}=\frac{\varphi_{i,r}f_{m}^{2}}{\psi_{i,r}^2-f_{m}^{2}+j\kappa_{i,r}f_{m}},\\\forall i\in\mathcal{N_C},\forall r\in\mathcal{R}, \forall m\in\mathcal{M},%, \angle\mathbf{\phi}_{i,r,m}=G(\theta_{i,r},f_m),\\
%\theta_{i,r}\in[-\pi,\pi], ~~\forall i\in\mathcal{N_C},.
\end{aligned}
\end{equation}
where $\varphi_{i,r}>0$, $\psi_{i,r}>0$, and $\kappa_{i,r}$ are the element-dependent oscillator strength, resonance frequency, and damping factor, respectively. These parameters can be tuned by external control for each element individually.
%\begin{subequations}\label{eq5}
%\begin{align}
%&F(\theta,f)=a_1G^2(\theta,f)+b_1G(\theta,f)+c_1,\\
%&G(\theta,f)=K(\theta)f+B(\theta),\\
%&K(\theta)=a_2sin(b_2\theta+c_2)+a_3sin(b_3\theta+c_3),\\
%&B(\theta)=a_4sin(b_4\theta+c_4)+a_5sin(b_5\theta+c_5),
%\end{align}
%\end{subequations}

For the MIMO-OFDM system with bandwidth $B$, the central frequency of each subcarrier $f_m$  is defined as $f_m=f_c+(m-\frac{M+1}{2})\frac{B}{M}, \forall m\in\mathcal{M}$.

After  CP is eliminated and  the discrete Fourier transform (DFT) is applied, the transmitted signal is converted back to the frequency domain symbol. Let $\mathbf{y}_{n_b,k,m}\in\mathbb{C}^{N_r}$  denote the baseband frequency domain signal on the tone $m$ from the BS $n_b$ to the user $k$. Based on the aforementioned channel model, $\mathbf{y}_{n_b,k,m}$  can be expressed as follows.
\begin{equation}\label{eq5}
    \begin{aligned}
         &\mathbf{y}_{n_b,k,m}=\mathbf{\hat{H}}_{n_b,k,m}^{H}\mathbf{x}_{n_b,m} \\
         & =\left(\mathbf{H}_{n_b,k,m}^{H}+\sum_{i=1}^{N_c}\mathbf{F}_{i,k,m}^{H}\mathbf{\Phi}_{i,m}^H\mathbf{G}_{n_b,i,m}\right) \sum_{j=1}^{K}\mathbf{w}_{n_b,m,j}s_{n_b,m,j}.
    \end{aligned}
\end{equation}

\subsection{Problem Fomulation}

Taking into account the transmit power  constraints of the BS and RC constraints at the IRS, we investigate the problem of maximizing WSR in the  IRS-assisted cell-free network presented. Define  $\mathbf{F}_{k,m}=[\mathbf{F}_{1,k,m}^T,\cdots,\mathbf{F}_{N_c,k,m}^T]^T$, $\mathbf{G}_{n_b,m}=[\mathbf{G}_{n_b,1,m}^T,\cdots,\mathbf{G}_{n_b,N_c,m}^T]^T$, $\mathbf{\hat{H}}_{k,m}=[\mathbf{\hat{H}}_{1,k,m}^T,\cdots,\mathbf{\hat{H}}_{N_b,k,m}^T]^T$, $\mathbf{\Phi}_{m}=diag(\mathbf{\Phi}_{1,m}\cdots,\mathbf{\Phi}_{N_c,m})$, and $\mathbf{w}_{m,k}=[\mathbf{w}_{1,m,k}^T,\cdots,\mathbf{w}_{N_b,m,k}^T]^T$, $\mathbf{s}_{m,k}=[{s}_{1,m,k},\cdots,{s}_{N_b,m,k}]^T$, the received signal $\mathbf{y}_{k,m}\in\mathbb{C}^{N_r}$ from the user $k$ on the tone $m$ is given by
\begin{equation}\label{eq6}
    \begin{aligned}
        &\mathbf{y}_{k,m}=\\
        &\sum_{n_b=1}^{N_b}\sum_{j=1}^{K}\left(\mathbf{H}_{n_b,m,k}^{H}\!+\mathbf{F}_{k,m}^{H}\mathbf{\Phi}_{m}^H\mathbf{G}_{n_b,m}\right)\mathbf{w}_{n_b,m,j}s_{n_b,m,j}\!+\mathbf{n}_{k,m} \\
        &=\sum_{j=1}^{K}\mathbf{\hat{H}}_{k,m}^H\mathbf{w}_{m,j}\mathbf{s}_{m,j}+\mathbf{n}_{k,m},
    \end{aligned}
\end{equation}
where $\mathbf{n}_{k,m}\triangleq[{n}_{k,m,1}^{T},\cdots,{n}_{k,m,N_r}^{T}]^{T}$  stands for  additive white Gauss noise (AWGN),  $\mathbf{n}_{k,m}\sim\mathcal{CN}(\mathbf{0}_{N_r},{\sigma}^2\mathbf{I}_{N_r}), \forall k\in\mathcal{K},\forall m\in\mathcal{M}$.

Based on the channel estimation techniques proposed in~\cite{IEEEconf:23},~\cite{IEEEconf:24}, we assume that all channel state information (CSI) can be accurately acquired in this study.
%All channel state information (CSI) can be accurately  achieved by employing the  channel estimation techniques~\cite{IEEEconf:24}-\cite{IEEEconf:29} is assumed in the paper, such as the brute-force method the semi-passive IRS scheme [24], the discrete Fourier transform matrix quantization~[26] and the compressed sensing method~[27], etc.

Then, the SINR of the transmit symbol $\mathbf{s}_{k,m}$ at the $k$-th user  on the $m$-th subcarrier is formulated as follows.
\begin{equation}\begin{split}\label{eq7}
\begin{aligned}
&\gamma_{k,m}= \\
&\mathbf{w}_{m,k}^H\mathbf{\hat{H}}_{k,m}\left(\sum_{j=1,j\neq k }^{K}\mathbf{\hat{H}}_{k,m}^H\mathbf{w}_{m,j}\left(\mathbf{\hat{H}}_{k,m}^{H}\mathbf{w}_{m,j}\right)^H+\mathbf{\Xi}_{k,m}\right)^{-1}\\
&\times\mathbf{\hat{H}}_{k,m}^{H}\mathbf{w}_{m,k},
\end{aligned}
\end{split}
\end{equation}
where $\mathbf{\Xi}_{k,m}={\sigma}^2\mathbf{I}_{N_r}$ is a  covariance matrix of $\mathbf{n}_{k,m}$.

The WSR for all $K$ users can thus be expressed  as follows.
\begin{equation}\label{eq8}
R_{sum}=\sum_{k=1}^{K}\sum_{m=1}^{M}\xi_{k,m}\log_2 \left(1+\gamma_{k,m}\right),
\end{equation}
where $\xi_{k,m}>0$  is the weight factor of the $k$-th user on the $m$-th subcarrier.%  and $R_{k,d}$  represents the rate of user $k$ on subcarrier $d$.

Hence, the WSR maximization optimization problem can  be expressed as
\begin{subequations}\label{eq9}
	\begin{align}
	\mathcal{P}(\mathrm{0}):~\underset{\mathbf{W}, \boldsymbol{\Phi}}\max\quad &R_{sum}(\mathbf{W}, \boldsymbol{\Phi})\!=\!{\sum_{k=1}^{K}\sum_{m=1}^{M}\xi_{k,m}\log_2 \left(1+\gamma_{k,m}\right)}\label{eq9a}\\
		\text{s.t.}\quad &\sum_{k=1}^{K}\sum_{m=1}^{M}||\mathbf{w}_{n_b,m,k}||^{2} \leq P_{\mathrm{n_b,max}},\forall n_b\in\mathcal{N_B},\label{eq9b} \\
                          &\mathbf{\phi}_{i,r,m} \in \Omega,\label{eq9c}\\
                          &|\mathbf{\phi}_{i,r,m}|\leq1,\label{eq9d}
                           \forall i\in\mathcal{N_C},\forall r\in\mathcal{R},\forall m\in\mathcal{M}.
	\end{align}
\end{subequations}
where $P_{\mathrm{n_b,max}}$  represents the maximum transmit power of the $n_b$-th BS, $\Omega$ is the set of all $\mathbf{\phi}_{i,r,m}$ satisfying equations (4).  $\boldsymbol{\Phi}\in\mathbb{C}^{MRN_c\times MRN_c}$ and $\mathbf{W}\in\mathbb{C}^{N_tN_bMK\times 1}$  is  defined as follows, respectively.
\begin{equation}\label{eq10}
\begin{aligned}
&\boldsymbol{\Phi}=diag(\boldsymbol{\Phi}_{1},\cdots,\boldsymbol{\Phi}_{2},\cdots,\boldsymbol{\Phi}_{M}),\\
&\mathbf{W}=\left[\mathbf{w}_{1,1}^T,\mathbf{w}_{1,2}^T,\cdots,\mathbf{w}_{1,K}^T,\mathbf{w}_{2,1}^T,\mathbf{w}_{2,2}^T,\cdots,\mathbf{w}_{M,K}^T\right]^{T}.
\end{aligned}
\end{equation}
The problem $\mathcal{P}(\mathrm{0})$ is a challenging high-dimensional non-convex optimization problem due to the deep coupling of variables in the objective function and the non-convex nature of the unit-modulus constraint.  In the next section, we employ the LDT and the MCQT from~\cite{IEEEconf:64}, [65] to decouple $\mathbf{W}$ and $\mathbf{\Phi}$. Subsequently, we propose the CADMM-APG-FRCG algorithm  to achieve an optimal solution to each subproblem within joint optimization.
%=======================================================section3=====================================================================
\section{PROPOSED SOLUTION}\label{sec:3}
In this section, we propose a simple and efficient approach to address the  WSR optimization problem $\mathcal{P}(\mathrm{0})$ in (9).
Firstly, we employ the LDT and MCQT from~\cite{IEEEconf:64}, [65] to convert $\mathcal{P}(\mathrm{0})$ into  $\mathcal{P}(\mathrm{1})$, thereby decoupling the variables
$\mathbf{W}$ and {$\mathbf{\Phi}$} in the objective function into  tractable subproblems  discussed in Subsection $\mathrm{III-A}$. Subsequently,  we present a  low-complexity algorithm to solve  joint precoding design problems concerning $\mathbf{W}$ and {$\mathbf{\Phi}$}, respectively, {which leverages the CADMM-APG-FRCG algorithm~\cite{IEEEconf:66}-\cite{IEEEconf:68}}.

%First, we transform  $\mathcal{P}(\mathrm{0})$  into $\mathcal{P}(\mathrm{1})$ using the Lagrangian dual transform in~\cite{IEEEconf:54} to decouple the variable $\mathbf{W}$ and $\mathbf{\Theta}$ in the objective function into  tractable subproblems in Subsection $\mathrm{III-A}$. Then, based on the  CADMM-APG algorithm~\cite{IEEEconf:55}-\cite{IEEEconf:57}, we develop a low complexity algorithm to solve the joint precoding design  problems in terms of $\mathbf{W}$ and $\mathbf{\Theta}$, respectively.% and then the $\mathcal{P}(\mathrm{1})$ problem is divided into three subproblems to solve alternately. The solutions of the three subproblems are given in Subsection III-B, III-C, and III-D respectively.

\subsection{Lagrangian Dual Transform}
To address the complex sum-of-logarithms-of-ratio problem in the WSR maximization problem $\mathcal{P}(\mathrm{0})$ in (9),  we employ the LDT  to decouple the logarithms. Let us introduce an auxiliary variable $\boldsymbol{\eta}\in\mathbb{R}^{MK}$ with $\boldsymbol{\eta}=\left[\eta_{1,1},\eta_{1,2},\cdots,\eta_{1,K},\eta_{2,1},\eta_{2,2},\cdots,\eta_{M,K}\right]^T$. Consequently, the problem $\mathcal{P}(\mathrm{0})$  is transformed into $\mathcal{P}(\mathrm{1})$.
\begin{equation}\label{eq11}
\begin{aligned}
\mathcal{P}(\mathrm{1}):~\underset{\mathbf{W}, {\boldsymbol{\Phi}},\boldsymbol{\eta}}\max \quad &f(\mathbf{W}, {\boldsymbol{\Phi}}, \boldsymbol{\eta}) \\
\text{s.t.} \quad & {(9b), (9c),(9d).}
%\sum_{k=1}^{K}\sum_{m=1}^{M}||\mathbf{w}_{n_b,m,k}||^{2} \leq P_{\mathrm{n_b,max}},\forall n_b\in\mathcal{N_B},\\
                         % &|\mathbf{\theta}_{i,r}|=1, \forall i\in\mathcal{N_C},\forall r\in\mathcal{R},
\end{aligned}
\end{equation}
where  $f(\mathbf{W}, {\boldsymbol{\Phi}}, \boldsymbol{\eta})$ is denoted by
\begin{equation}\label{eq12}
\begin{aligned}
&f(\mathbf{W}, {\boldsymbol{\Phi}}, \boldsymbol{\eta})=\sum_{k=1}^{K}\sum_{m=1}^{M}{\xi_{k,m}}\log_2\left(1+\eta_{k,m}\right)\\
&-\sum_{k=1}^{K}\sum_{m=1}^{M}{\xi_{k,m}}\eta_{k,m}
\!+\!{\sum_{k=1}^{K}\sum_{m=1}^{M}} {\xi_{k,m}}\left(1\!+\!\eta_{k,m}\right)f_{k,m}(\mathbf{W}, {\boldsymbol{\Phi}}),
\end{aligned}
\end{equation}
where  $f_{k,m}(\mathbf{W},{\boldsymbol{\Phi}})$  is defined as
\begin{equation}\label{eq13}
\begin{aligned}
&f_{k,m}(\mathbf{W}, {\boldsymbol{\Phi}})=\\
&\mathbf{w}_{m,k}^H{\mathbf{\hat{H}}_{k,m}}\left(\sum_{j=1}^{K}{\mathbf{\hat{H}}_{k,m}^H}\mathbf{w}_{m,j}\left({\mathbf{\hat{H}}_{k,m}^H}\mathbf{w}_{m,j}\right)^H+\mathbf{\Xi}_{k,m}\right)^{-1}\\
&\times{\mathbf{\hat{H}}_{k,m}^H}\mathbf{w}_{m,k},
\end{aligned}
\end{equation}
Then, the {CADMM-APG-FRCG} algorithm is devised to alternatively solve the variables $\boldsymbol{\eta}$, $\mathbf{W}$ and {$\boldsymbol{\Phi}$}.% in an alternating manner.
\subsection{Fix $\left(\mathbf{W},{\mathbf{\boldsymbol{\Phi}}}\right)$  and solve $\boldsymbol{\eta}^\mathrm{opt}$ }
When $\left(\mathbf{W}^*,{\mathbf{\boldsymbol{\Phi}}^*}\right)$ remains unchanged, the optimal solution $\boldsymbol{\eta}$ in (12) is obtained by setting $\partial f/\partial\eta_{k,m}=0$.% which is given by
\begin{equation}\label{eq14}
\begin{aligned}
\eta_{k,m}^\mathrm{opt}=\gamma_{k,m}^*,\forall k\in\mathcal{K},\forall m\in\mathcal{M}.
\end{aligned}
\end{equation}
By substituting $\eta_{k,m}^\mathrm{opt}=\gamma_{k,m}^*$ (14) into (12), we can observe that only the final term in (12)  depends on both variables {$\mathbf{\boldsymbol{\Phi}}$} and $\mathbf{W}$. Thus, we can simplify (11) as  optimization problems for {$\mathbf{\boldsymbol{\Phi}}$} and $\mathbf{W}$. %The next two subsections will address these sub-problems.
\subsection{Fix $\left({\mathbf{\boldsymbol{\Phi}}},\boldsymbol{\eta}\right)$  and solve $\mathbf{W}^\mathrm{opt}$ }
Given $\left({\mathbf{\boldsymbol{\Phi}}^*},\boldsymbol{\eta}^*\right)$, $\mathcal{P}(\mathrm{1})$ is equivalent to the following  $\mathcal{P}(\mathrm{2})$.
\begin{equation}\label{eq15}
\begin{aligned}
\mathcal{P}(\mathrm{2}):~\underset{\mathbf{W}}\max \quad &f_1(\mathbf{W})=\sum_{k=1}^{K}\sum_{m=1}^{M}\zeta_{k,m}f_{k,m}(\boldsymbol{\mathbf{W},{\Phi}^{*}})\\
\text{s.t.} \quad & \sum_{k=1}^{K}\sum_{m=1}^{M}||\mathbf{w}_{n_b,m,k}||^{2} \leq P_{\mathrm{n_b,max}},\forall n_b\in\mathcal{N_B},
\end{aligned}
\end{equation}
where $\zeta_{k,m}={\xi_{k,m}}\left(1+\eta_{k,m}^*\right)$. Since $f_{k,m}(\mathbf{W}, {\boldsymbol{\Phi}^{*}})$ in (15) represents a complex sum-of-fractions problem that is non-convex, the direct solution of (15) becomes highly challenging. To address this issue,  we employ  the MCQT, as presented in~\cite{IEEEconf:64}, by introducing auxiliary variables $\boldsymbol{\delta}_{m,k}\in\mathbb{C}^{N_r}$ with $\boldsymbol{\delta}=\left[\boldsymbol{\delta}_{1,1},\boldsymbol{\delta}_{1,2},\cdots,\boldsymbol{\delta}_{1,K},\boldsymbol{\delta}_{2,1},\boldsymbol{\delta}_{2,2},
\cdots,\boldsymbol{\delta}_{M,K}\right]$. Consequently, the $\mathcal{P}(\mathrm{2})$  is reformulated as
\begin{equation}\label{eq16}
\begin{aligned}
\underset{\mathbf{W},\boldsymbol{\delta}}\max~~&f_2(\mathbf{W},\boldsymbol{\delta})=\sum_{k=1}^{K}\sum_{m=1}^{M}2\sqrt{\zeta_{k,m}}\Re\{  \boldsymbol{\delta}_{k,m}^H{\boldsymbol{\hat{H}}_{k,m}^H}\boldsymbol{w}_{m,k}\}\\
&-\sum_{k=1}^{K}\sum_{m=1}^{M}\boldsymbol{\delta}_{k,m}^H\\
&\times\left(\sum_{j=1}^{K}{\mathbf{\hat{H}}_{k,m}^H}\mathbf{w}_{m,j}\left({\mathbf{\hat{H}}_{k,m}^H}\mathbf{w}_{m,j}\right)^H+\mathbf{\Xi}_{k,m}\right)\boldsymbol{\delta}_{k,m}\\
\text{s.t.} \quad & \sum_{k=1}^{K}\sum_{m=1}^{M}||\mathbf{w}_{n_b,m,k}||^{2} \leq P_{\mathrm{n_b,max}},\forall n_b\in\mathcal{N_B}.
\end{aligned}
\end{equation}
According to (16), solving  $\mathbf{W}$ in $\mathcal{P}(\mathrm{2})$ is equivalent to an alternative updating of  $\boldsymbol{\delta}$  and $\mathbf{W}$ in (16), which will be addressed separately below.

\textit{C1}.~\textit{Fix $\mathbf{W}$ and Solve $\boldsymbol{\delta}^\mathrm{opt}$}
\

Fixing $\mathbf{W}$ in $f_2$,  and setting $\partial f_2/\partial\boldsymbol{\delta}_{k,m}=0$, we can obtain the optimal solution for $\boldsymbol{\delta}$.
\begin{equation}\label{eq17}
\begin{aligned}
\boldsymbol{\delta}_{k,m}^\mathrm{opt}=&
\sqrt{\zeta_{k,m}}\!{\left(\sum_{j=1 }^{K}{\mathbf{\hat{H}}_{k,m}^H}\mathbf{w}_{m,j}{\left({\mathbf{\hat{H}}_{k,m}^H}\mathbf{w}_{m,j}\right)^H}+\mathbf{\Xi}_{k,m}\right)}^{-1}\\
&\times{{\mathbf{\hat{H}}_{k,m}^H}\mathbf{w}_{m,k}}.
\end{aligned}
\end{equation}

\textit{C2}.~\textit{Fix $\boldsymbol{\delta}$ and Solve $\mathbf{W}^\mathrm{opt}$}
\

The optimal value  $\boldsymbol{\delta}$ obtained in (17) is substituted into $f_2$, resulting in $f_2$ being solely associated with the variable $\mathbf{W}$. Define
\begin{subequations}\label{eq18}
\begin{align}
&\mathbf{d}_m=\sum_{k=1 }^{K}{\mathbf{\hat{H}}_{k,m}}\boldsymbol{\delta}_{k,m}\boldsymbol{\delta}_{k,m}^H{\mathbf{\hat{H}}_{k,m}^H}\label{eq18a},\\
&\mathbf{D}_m=\mathbf{I}_K\otimes\mathbf{d}_m,\mathbf{c}_{k.m}={\mathbf{\hat{H}}_{k,m}}\boldsymbol{\delta}_{k,m}.\label{eq18b}
\end{align}
\end{subequations}
Then $f_2$  can be reformulated as
\begin{equation}\label{eq19}
\begin{aligned}
&f_2(\mathbf{W})=-\mathbf{W}^H\mathbf{D}\mathbf{W}+\Re\{2\mathbf{C}^H\mathbf{W}\}-U,
\end{aligned}
\end{equation}
where
\begin{subequations}\label{eq20}
\begin{align}
&\mathbf{D}=diag\left (\mathbf{D}_1,\cdots,\mathbf{D}_M \right ),U=\sum_{k=1}^{K}\sum_{m=1}^{M}\boldsymbol{\delta}_{k,m}^H\mathbf{\Xi}_{k,m}\boldsymbol{\delta}_{k,m}\label{eq20a},\\
&\mathbf{C}=\left[\mathbf{c}_{1,1}^T,\mathbf{c}_{1,2}^T,\cdots,\mathbf{c}_{1,K}^T,\mathbf{c}_{2,1}^T,\mathbf{c}_{2,2}^T,\cdots,\mathbf{c}_{M,K}^T\right]^{T}.\label{eq20b}
\end{align}
\end{subequations}
The $f_2$ in (16) can be simplified further.
\begin{equation}\label{eq21}
\begin{aligned}
\mathcal{P}(\mathrm{3}):~\underset{\mathbf{W}}\min \quad &f_3(\mathbf{W})=\mathbf{W}^H\mathbf{D}\mathbf{W}-\Re\{2\mathbf{C}^H\mathbf{W}\}\\
\text{s.t.} \quad & \mathbf{W}^H\mathbf{P}_{n_b}\mathbf{W} \leq P_{\mathrm{n_b,max}},\forall n_b\in\mathcal{N_B},
\end{aligned}
\end{equation}
where $\mathbf{P}_{n_b}=\mathbf{I}_{MK}\otimes\{(\mathbf{e}_{n_b}\mathbf{e}_{n_b}^H)\otimes\mathbf{I}_{N_t}\},\mathbf{e}_{n_b}\in\mathbb{R}^{N_b}$.  The matrices $\mathbf{D}$ and $\mathbf{P}_{n_b},\forall {n_b}\in \mathcal{N_B}$ being positive semidefinite, $\mathcal{P}(\mathrm{3})$ in (21)  can be classified as a standard QCQP problem, which is generally known to be  NP-hard. We employ the CADMM to optimally solve the QCQP.

\textit{C3}.~\textit{CADMM algorithm to solve $\mathbf{W}$}
\

We employ CADMM to solve the  subproblem of active beamforming (21).  CADMM  is derived from  ADMM~\cite{IEEEconf:69},~\cite{IEEEconf:70}. To reduce  computational complexity and address the challenging subproblems of ADMM, a consistency strategy known as CADMM is utilized~\cite{IEEEconf:66}. Both ADMM and CADMM have been extensively applied in addressing  convex and non-convex optimization problems~\cite{IEEEconf:66},~\cite{IEEEconf:69},~\cite{IEEEconf:70}. For a complete understanding and practical application of  ADMM and CADMM,  readers can refer to the literature~\cite{IEEEconf:66},~\cite{IEEEconf:69},~\cite{IEEEconf:70}.

Let $\mathbf{V}_{n_b}=\mathbf{W}$, and $\mathbf{V}_{n_b}$ be an auxiliary variable. Then $\mathcal{P}(\mathrm{3})$ can be equivalently formulated as follows.
\begin{subequations}\label{eq22}
\begin{align}
\nonumber\mathcal{P}(\mathrm{3'}):~\underset{\mathbf{W}}\min \quad &f_3(\mathbf{W})\\
\text{s.t.} \quad &\mathbf{W=V}_{n_b}\label{eq22a},\\
&\mathbf{V}_{n_b}^H\mathbf{P}_{n_b}\mathbf{V}_{n_b} \leq P_{\mathrm{n_b,max}},\forall n_b\in\mathcal{N_B}\label{eq22b}.
\end{align}
\end{subequations}
 The augmented Lagrangian function of $\mathcal{P}(\mathrm{3'})$ is %Introducing a penalty parameter $\mathbf{\rho}$ ,
\begin{equation}\label{eq23}
\begin{aligned}
L_\alpha(\mathbf{W},\mathbf{V}_{n_b},\mathbf{q}_{n_b})=&\mathbf{W}^H\mathbf{D}\mathbf{W}-\Re\{2\mathbf{C}^H\mathbf{W}\}\\&+\alpha\sum_{n_b=1 }^{N_b}||\mathbf{W}-\mathbf{V}_{n_b}+\mathbf{q}_{n_b}||^2,\\
\end{aligned}
\end{equation}
where $\mathbf{\alpha}>0$ is a penalty parameter. $\mathbf{q}_{n_b}$%=[\mathbf{u}_1,\cdots,\mathbf{u}_M]$
~represents the Lagrange dual vector. The CADMM method is employed to solve ${P}(\mathrm{3'})$, which follows an iterative form.
\begin{subequations}\label{eq24}
\begin{align}
\mathbf{W}^{j+1}&=\arg\underset{\mathbf{W}}\min \quad L_\alpha\left(\mathbf{W},\mathbf{V}_{n_b}^j,\mathbf{q}_{n_b}^j\right)\label{eq24a},\\
\mathbf{V}_{n_b}^{j+1}&=\arg\underset{\mathbf{V}}\min \quad L_\alpha\left(\mathbf{W}^j,\mathbf{V}_{n_b},\mathbf{q}_{n_b}^j\right)\label{eq24b},\\
\mathbf{q}_{n_b}^{j+1}&=\mathbf{q}_{n_b}^j+\mathbf{W}^{j+1}-\mathbf{V}_{n_b}^{j+1}\label{eq24c}, \forall n_b\in\mathcal{N_B}.
\end{align}
\end{subequations}

\subsubsection{Solving $\mathbf{W}$ in (24a)}%$C3.~1)$
\

To avoid the computational burden associated with directly solving (24a), which requires inverting a high-dimensional matrix $\mathbf{D}$ of size $N_bN_tMK$, we employ a linearization strategy to address the quadratic term $\mathbf{W}^H\mathbf{D}\mathbf{W}$.

%The direct solution of (24a) involves the computation of the inversion of  matrix $\mathbf{D}$, which has a high-dimension $N_bN_tZK$. Due to the computational complexity associated with  calculating the inverse matrix, it is necessary to avoid this issue when dealing with large-scale optimization problems. To obtain an efficient algorithm without computing the inverse matrix, we employ a consistency strategy for handling $\mathbf{W}^H\mathbf{D}\mathbf{W}$.

Specifically, we perform a first-order Taylor expansion of $\mathbf{W}^H\mathbf{D}\mathbf{W}$ at any given point $\mathbf{W}_0$, then
\begin{equation}\label{eq25}
\begin{aligned}
\mathbf{W}^H\mathbf{D}\mathbf{W}=&
\mathbf{W}_0^H\mathbf{D}\mathbf{W}_0+\Re\{2\mathbf{D}\mathbf{W}_0\left( \mathbf{W}-\mathbf{W}_0 \right)\}\\&
+\beta ||\mathbf{W}-\mathbf{W}_0||^2,
\end{aligned}
\end{equation}
where $\beta>0$ is a penalty parameter.  (24a)  is equivalent to solving the following problem.
\begin{equation}\label{eq26}
\begin{aligned}
\underset{\mathbf{W}}\min \quad &\Re\{2\mathbf{D}\mathbf{W}_0\left( \mathbf{W}-\mathbf{W}_0 \right)\}+\beta ||\mathbf{W}-\mathbf{W}_0||^2\\&
-\Re\{2\mathbf{C}^H\mathbf{W}\}+\alpha\sum_{n_b=1 }^{N_b}||\mathbf{W}-\mathbf{V}_{n_b}^{j}+\mathbf{q}_{n_b}^{j}||^2
\end{aligned}
\end{equation}
%Setting the derivative of the objective function on $\mathbf{W}$ equals 0, we get
The derivative of the objective function with respect to $\mathbf{W}$ is set to zero, yielding
\begin{equation}\label{eq27}
\begin{aligned}
&\mathbf{W}^{j+1}\!=\frac{1}{N_b\alpha+\beta}\left[\beta\mathbf{W}_0\!+\mathbf{C}\!-\mathbf{D}\mathbf{W}_0\!+\alpha\sum_{n_b=1 }^{N_b}\left(\mathbf{V}_{n_b}^{j}\!-\mathbf{q}_{n_b}^{j}\right)\right].
\end{aligned}
\end{equation}

\subsubsection{Solving $\mathbf{V}_{n_b}$ in (24b)}% $C3.~2)$
\

(24b) requires one to address the following optimization issue.
\begin{equation}\label{eq28}
\begin{aligned}
\underset{\mathbf{V}_{n_b}}\min \quad &\sum_{{n_b}=1}^{N_b}||\mathbf{W}^{j+1}-\mathbf{V}_{n_b}+\mathbf{q}_{n_b}^j||^2\\
\text{s.t.} \quad &\mathbf{V}_{n_b}^H\mathbf{P}_{n_b}\mathbf{V}_{n_b} \leq P_{\mathrm{n_b,max}}, \forall n_b\in\mathcal{N_B}.
\end{aligned}
\end{equation}
Problem (28) is a QCQP with $N_b$ constraints, which  is generally known to be NP-hard. However, considering the special characteristics of this problem, we can equivalently transform it into $N_b$ QCQPs  with only one constraint each (QCQP-1), thereby enabling an optimal solution.

%(28) is equivalent to  $N_b$ QCQP with one constraint only (QCQP-1), thus can be solved  optimally.

%We note the characteristic that, for a vector, we can reorder its elements without changing the value of the 2-norm of the vector. Hence, we reorder the elements of the variables $\mathbf{W}$, $\mathbf{V}_{n_b}$ and $\mathbf{q}_{n_b}$. Specifically, let $\mathbf{W}^{j+1}=[\mathbf{W}_{1}^{j+1},\cdots,\mathbf{W}_{N_b}^{j+1}]$,  which  reorders  $\mathbf{W}$ in (10), $\mathbf{V}=[\mathbf{V}_{1},\cdots,\mathbf{V}_{n_b}]$, $\mathbf{q}^{j+1}=[\mathbf{q}_{1}^{j+1},\cdots,\mathbf{q}_{n_b}^{j+1}]$. Here, $\mathbf{W}_{n_b}^{j+1}=\left[{\mathbf{w}_{n_b,1,1}^{j+1}}^T,\cdots,{\mathbf{w}_{n_b,1,K}^{j+1}}^T,\cdots,{\mathbf{w}_{n_b,2,K}^{j+1}}^T,
%\cdots,{\mathbf{w}_{n_b,Z,K}^{j+1}}^T\right], \\\forall n_b\in\mathcal{N_B}$. The definitions of $\mathbf{V}_{n_b}$ and $\mathbf{q}_{n_b}^{j+1}$ are similar to that of $\mathbf{W}^{j+1}$.  Considering that $\mathbf{P}_{n_b}$ is a 0-1 matrix,
Based on the definition of $\mathbf{P}_{n_b}$, which is a 0-1 matrix, and (28) is equivalent to the following $N_b$ QCQP-1.
\begin{equation}\label{eq29}
\begin{aligned}
\underset{\mathbf{V}_{n_b}}\min \quad &||\mathbf{W}^{j+1}-\mathbf{V}_{n_b}+\mathbf{q}_{n_b}^j||^2\\
\text{s.t.} \quad &\mathbf{V}_{n_b}^H\mathbf{V}_{n_b} \leq P_{\mathrm{n_b,max}}.
\end{aligned}
\end{equation}
Let $\boldsymbol{\epsilon}_{n_b}^j=\mathbf{W}^{j+1}+\mathbf{q}_{n_b}^j$, %then, (29) can be decomposed into $N_b$ subproblems
%\begin{equation}\label{eq30}
%\begin{aligned}
%\underset{\mathbf{V}_{n_b}}\min \quad &||\mathbf{V}_{n_b}-\boldsymbol{\epsilon}_{n_b}^j||^2\\
%\text{s.t.} \quad &\mathbf{V}_{n_b}^H\mathbf{P}_{n_b}\mathbf{V}_{n_b} \leq P_{\mathrm{n_b,max}}.%\forall m\in\mathcal{M},
%\end{aligned}
%\end{equation}
the Lagrange function of (29) is
\begin{equation}\label{eq30}
L_{\sigma_{n_b}}(\mathbf{V}_{n_b},\sigma_{n_b})=||\mathbf{V}_{n_b}-\boldsymbol{\epsilon}_{n_b}^j||^2+\sigma_{n_b}(\mathbf{V}_{n_b}^H\mathbf{V}_{n_b}-P_{\mathrm{n_b,max}}),\\
\end{equation}
where $\sigma_{n_b}$ is a Lagrange multiplier. Setting $\frac{\partial L_\sigma(\mathbf{V}_{n_b},\boldsymbol{\epsilon}_{n_b}^j)}{\partial \mathbf{V}_{n_b}}=0$, we get the optimal solution as follows.
\begin{equation}\label{eq31}
\mathbf{V}_{n_b}^{j+1}(\sigma_{n_b})=\frac{1}{(1+\sigma_{n_b})}\boldsymbol{\epsilon}_{n_b}^j.
\end{equation}
The optimal dual variable $\sigma_{n_b}$ should satisfy the following complementary slackness condition.
\begin{equation}\label{eq32}
\sigma_{n_b}(\mathbf{V}_{n_b}^H\mathbf{V}_{n_b}-P_{\mathrm{n_b,max}})=0.
\end{equation}
If $\mathbf{V}_{n_b}^H\mathbf{V}_{n_b}<P_{\mathrm{n_b,max}}$, then $\sigma_{n_b}=0$. Otherwise, we need to solve the equation
$\mathbf{V}_{n_b}^H\mathbf{V}_{n_b}=P_{\mathrm{n_b,max}}$. By substituting (31) into the  given equality, we derive the subsequent equation.
\begin{equation}\label{eq33}
\frac{1}{(1+\sigma_{n_b})^2}||{\boldsymbol{\epsilon}_{n_b}^j}||^2=P_{\mathrm{n_b,max}}.
%{\boldsymbol{\varpi}_{m}^j}^H(\mathbf{I}+\lambda\mathbf{D}_m)^{-1}\mathbf{D}_m(\mathbf{I}+\lambda\mathbf{D}_m)^{-1}\boldsymbol{\varpi}_{m}^j=P_{\mathrm{m,max}}
\end{equation}
Hence,
\begin{equation}\label{eq34}
\sigma_{n_b}=\frac{||{\boldsymbol{\epsilon}_{n_b}^j}||}{\sqrt{P_{\mathrm{n_b,max}}}}-1.
%\sum_{k=1}^{K}\sum_{d=1}^{D}\frac{1}{(1+\lambda_{m})^2}|\boldsymbol{\varpi}_{k,d}^j|^2=P_{\mathrm{m,max}}
\end{equation}
Plugging (34) into (31), we get the optimal solution of $\mathbf{V}_{n_b}$.
\begin{equation}\label{eq35}
\mathbf{V}_{n_b}^{j+1}=\frac{\sqrt{P_{\mathrm{n_b,max}}}}{||{\boldsymbol{\epsilon}_{n_b}^j}||}\boldsymbol{\epsilon}_{n_b}^j
=\sqrt{P_{\mathrm{n_b,max}}}e^{j\boldsymbol{\vartheta}(\boldsymbol{\epsilon}_{n_b}^j)}.
\end{equation}
where $\boldsymbol{\vartheta}(\boldsymbol{\epsilon}_{n_b}^j)$ denotes the  phase angle of $\boldsymbol{\epsilon}_{n_b}^j$. Therefore, we acquire  $\mathbf{V}^{j+1}=[\mathbf{V}_{1}^{j+1},\cdots,\mathbf{V}_{N_b}^{j+1}]$ in (28) by parallel computation.

Remark 1: the precoding vector $\mathbf{W}$ and the Lagrange multipliers $\boldsymbol{\sigma}=[\sigma_1,\cdots,\sigma_{N_b}]^{T}$ were obtained using the primal-dual subgradient  method (PDS) as described in~\cite{IEEEconf:38}. In contrast to the PDS method, which involves iteratively solving a high-dimensional optimization problem for updating $\mathbf{W}$ and $\boldsymbol{\sigma}$, our proposed approach is characterized by its simplicity and efficiency. This is because each subproblem in (24a) and (24b) has a closed-form solution that does not require iterative solving. Algorithm 1 summarizes the CADMM algorithm employed to solve the active precoding $\mathbf{W}$.

%The CADMM algorithm for solving the active precoding $\mathbf{W}$ is summarized in Algorithm 1.
\begin{algorithm}
	\caption{CADMM algorithm to solve the active precoding $\mathbf{W}$}
	\begin{algorithmic}
		%\STATE {1: $\mathbf{Input}$: $\mathbf{\delta}$, ${\mathbf{\xi}}$.}
		\STATE {1: Given $\left(\mathbf{\boldsymbol{\Phi}},\boldsymbol{\eta}\right)$}.
		\STATE {2: $\mathbf{Repeat}$}
		%\STATE {3: Update $\boldsymbol{\alpha}$ by (\ref{eq14});}
		\STATE {3: Update $\boldsymbol{\delta}$ by (\ref{eq17});}
		\STATE {4: Update $\mathbf{W}$ by (\ref{eq27});}
        \STATE {5: Update $\mathbf{V}$ by (\ref{eq35});}
        \STATE {6: Update $\mathbf{q}$ by (24c);}
		\STATE {7: $\mathbf{Until}$  the problem $\mathcal{P}(2)$ converges.}
	\end{algorithmic}
\end{algorithm}

\subsection{Fix $\left(\boldsymbol{\eta},\mathbf{W}\right)$  and solve ${\mathbf{\boldsymbol{\Phi}}^\mathrm{opt}}$ }
Given $\boldsymbol{\eta}$ and $\mathbf{W}$, we optimize ${\mathbf{\boldsymbol{\Phi}}}$ to address the $\mathcal{P}(\mathrm{1})$ problem in a more refined manner.
\begin{equation}\label{eq36}
\begin{aligned}
\mathcal{P}(\mathrm{4}):~\underset{{\boldsymbol{\Phi}}}\max \quad &f_4({\boldsymbol{\Phi}})=\sum_{k=1}^{K}\sum_{m=1}^{M}\zeta_{k,m}f_{k,m}({\boldsymbol{\Phi}},\mathbf{W}^*)\\
\text{s.t.}~~&{(9c), (9d).}%\mathbf{\phi}_{i,r,m}\in \Omega, \forall i\in\mathcal{N_C},\forall r\in\mathcal{R},\forall m\in\mathcal{M}.%\quad &\boldsymbol{\theta}^H\mathbf{E}_{m}\boldsymbol{\theta}=1, \forall m\in\mathcal{M},
\end{aligned}
\end{equation}
%where $\mathbf{E}_{m}=(\mathbf{e}_{r}\mathbf{e}_{r}^H)\otimes\mathbf{I}_{N_c},\mathbf{e}_{r}\in\mathbb{R}^{N_c},\forall m\in\mathcal{M},\mathcal{M}=\{1,\cdots,N_cR\}$.
Introduce a novel auxiliary function as
\begin{equation}\label{eq37}
\mathbf{T}_{k,m,j}({\mathbf{\Phi}_{m}})=\sum_{n_b=1}^{N_b}\left(\mathbf{H}_{n_b,k,m}^{H}+\mathbf{F}_{k,m}^{H}{\mathbf{\Phi}_{m}^H}\mathbf{G}_{n_b,m}\right)\mathbf{w}_{n_b,m,j}.
\end{equation}
$f_4$ can be rewritten as
\begin{equation}\label{eq38}
\begin{aligned}
&f_4({\boldsymbol{\Phi}})=\sum_{k=1}^{K}\sum_{m=1}^{M}\sqrt{\zeta_{k,m}}\mathbf{T}_{k,m,k}^H({\mathbf{\Phi}_{m}})\\
&\left(\sum_{j=1}^{K}\mathbf{T}_{k,m,j}({\mathbf{\Phi}_{m}})
\mathbf{T}_{k,m,j}^H({\mathbf{\Phi}_{m}})+\mathbf{\Xi}_{k,m}\right)^{-1}\mathbf{T}_{k,m,k}({\mathbf{\Phi}_{m}}).
\end{aligned}
\end{equation}
Let $\boldsymbol{\rho}=\left[\boldsymbol{\rho}_{1,1},\boldsymbol{\rho}_{1,2},\cdots,\boldsymbol{\rho}_{1,K},\boldsymbol{\rho}_{2,1},
\boldsymbol{\rho}_{2,2},\cdots,\boldsymbol{\rho}_{M,K}\right]$, where $\boldsymbol{\rho}_{m,k}\in\mathbb{C}^{N_r}$ is an auxiliary variable.  Employing the MCQT in [64], the problem $\mathcal{P}(\mathrm{4})$ is  reformulated as
\begin{equation}\label{eq39}
\begin{aligned}
\mathcal{P}(\mathrm{5}):~\underset{{\boldsymbol{\Phi}}}\max \quad &f_5({\boldsymbol{\Phi}},\boldsymbol{\rho})=\sum_{k=1}^{K}\sum_{m=1}^{M}g_{k,m}({\mathbf{\Phi}_{m}},\boldsymbol{\rho})\\
\text{s.t.}~~ &{(9c),(9d),}%\mathbf{\phi}_{i,r,m}\in \Omega, \forall i\in\mathcal{N_C},\forall r\in\mathcal{R},\forall m\in\mathcal{M},
\end{aligned}
\end{equation}
where
\begin{equation}\label{eq40}
\begin{aligned}
&g_{k,m}({\mathbf{\Phi}_{m}},\boldsymbol{\rho})=2\sqrt{\zeta_{k,m}}\Re\{ \boldsymbol{\rho}_{k,m}^H\mathbf{T}_{k,m,k}({\mathbf{\Phi}_{m}})\}\\
&-\boldsymbol{\rho}_{k,m}^H\left(\sum_{j=1}^{K}
\mathbf{T}_{k,m,j}({\mathbf{\Phi}_{m}})\}\mathbf{T}_{k,m,j}^{H}({\mathbf{\Phi}_{m}})+\mathbf{\Xi}_{k,m}\right)\boldsymbol{\rho}_{k,m}.\\
\end{aligned}
\end{equation}
The variables $\boldsymbol{\rho}$ and ${\boldsymbol{\Phi}}$ in the problem $\mathcal{P}(\mathrm{5})$ are  updated alternatively. The update steps for the variables $\boldsymbol{\rho}$ and $\boldsymbol{\Phi}$ are executed as follows.

\textit{D1}.~\textit{Fix ${\boldsymbol{\Phi}}$ and Solve $\boldsymbol{\rho}^\mathrm{opt}$}
\

The optimal value of $\boldsymbol{\rho}$ is obtained by fixing ${\boldsymbol{\Phi}}$  in $f_5$  and setting $\partial f_5/\partial\boldsymbol{\rho}_{k,m}=0$.
\begin{equation}\label{eq41}
\begin{aligned}
&\boldsymbol{\rho}_{k,m}^\mathrm{opt}=\sqrt{\zeta_{k,m}}\\
&\left(\sum_{j=1}^{K}\mathbf{T}_{k,m,j}({\mathbf{\Phi}_{m}})
\mathbf{T}_{k,m,j}^H({\mathbf{\Phi}_{m}})+\mathbf{\Xi}_{k,m}\right)^{-1}\mathbf{T}_{k,m,k}({\mathbf{\Phi}_{m}}).
\end{aligned}
\end{equation}

\textit{D2}.~\textit{Fix $\boldsymbol{\rho}$ and Solve ${\boldsymbol{\Phi}^\mathrm{opt}}$}
\

By substituting   the optimal value  $\boldsymbol{\rho}$  obtained in (41) into $f_5$, it is observed that   $f_5$  only depends on the variable ${\boldsymbol{\Phi}}$. Utilizing the  function $\mathbf{T}_{k,m,j}({\mathbf{\Phi}_{m}})$  from (38) and ${\mathbf{\Phi}_{m}}=diag({\mathbf{\Phi}_{1,m}}\cdots,{\mathbf{\Phi}_{N_c,m}})$, we can simplify the expression related to $f_5$.
\begin{equation}\label{eq42}
\begin{aligned}
&\boldsymbol{\rho}_{k,m}^H\mathbf{T}_{k,m,j}({\mathbf{\Phi}_{m}})\\&
\!{=\sum_{n_b=1}^{N_b}}\left(\boldsymbol{\rho}_{k,m}^H\mathbf{H}_{n_b,k,m}^{H}\mathbf{w}_{n_b,m,j}
\!+\boldsymbol{\rho}_{k,m}^H\mathbf{F}_{k,m}^{H}{\mathbf{\Phi}_{m}^H}\mathbf{G}_{n_b,m}\mathbf{w}_{n_b,m,j}\right)\\&
=\sum_{n_b=1}^{N_b}\boldsymbol{\rho}_{k,m}^H\mathbf{H}_{n_b,k,m}^{H}\mathbf{w}_{n_b,m,j}\\&
+{\boldsymbol{\phi}_{m}^{H}}\sum_{n_b=1}^{N_b}diag\left(\boldsymbol{\rho}_{k,m}^H
\mathbf{F}_{k,m}^{H}\right)\mathbf{G}_{n_b,m}\mathbf{w}_{n_b,m,j}\\&
=b_{k,m,j}+{\boldsymbol{\phi}_{m}^{H}}\boldsymbol{\varpi}_{k,m,j},
\end{aligned}
\end{equation}
where
\begin{subequations}\label{eq43}
\begin{align}
&{\boldsymbol{\phi}_{m}=\mathbf{\Phi}_{m}\mathbf{1}_{RN_{c}},}\\%[\mathbf{\Phi}_{1,m}\mathbf{1}_{R},\cdots,\mathbf{\Phi}_{N_c,m}\mathbf{1}_{R}]^{T}\label{eq43a},\\
&b_{k,m,j}=\sum_{n_b=1}^{N_b}\boldsymbol{\rho}_{k,m}^H\mathbf{H}_{n_b,k,m}^{H}\mathbf{w}_{n_b,m,j}\label{eq43b},\\
&\boldsymbol{\varpi}_{k,m,j}=\sum_{n_b=1}^{N_b}diag\left(\boldsymbol{\rho}_{k,m}^H
\mathbf{F}_{k,m}^{H}\right)\mathbf{G}_{n_b,m}\mathbf{w}_{n_b,m,j}.\label{eq43c}
\end{align}
\end{subequations}
Plugging  (42) into (40), $g_{k,m}$ can be rewritten as
\begin{equation}\label{eq44}
\begin{aligned}
&g_{k,m}({\boldsymbol{\phi}_{m}})=2\sqrt{\zeta_{k,m}}\Re\{b_{k,m,k}+{\boldsymbol{\phi}_{m}^H}\boldsymbol{\varpi}_{k,m,k}\}\\
&-\sum_{j=1}^{K}\left(b_{k,m,j}+{\boldsymbol{\phi}_{m}^H}\boldsymbol{\varpi}_{k,m,j}\right)\left(b_{k,m,j}^*+
\boldsymbol{\varpi}_{k,m,j}^H{\boldsymbol{\phi}_{m}}\right)\\
&-\boldsymbol{\rho}_{k,m}^H\boldsymbol{\Xi}_{k,m}\boldsymbol{\rho}_{k,m}.\\
\end{aligned}
\end{equation}
Then $f_5$ in (39) can be refined in the following manner.

{\begin{equation}\label{eq45}
\begin{aligned}
f_5(\boldsymbol{\phi}_{m})=\sum_{m=1}^{M}\left(-\boldsymbol{\phi}_{m}^H\boldsymbol{Q}_{m}\boldsymbol{\phi}_{m}
+2\Re\{\boldsymbol{\phi}_{m}^H\boldsymbol{\upsilon}_{m}\}\right)-P,\\
\end{aligned}
\end{equation}}
where
\begin{subequations}\label{eq46}
\begin{align}
&{\boldsymbol{Q}_m=\sum_{k=1}^{K}\sum_{j=1}^{K}\boldsymbol{\varpi}_{k,m,j}\boldsymbol{\varpi}_{k,m,j}^H\label{eq46a},}\\
&{\boldsymbol{\upsilon}_m=\sum_{k=1}^{K}\sqrt{\zeta_{k,m}}\boldsymbol{\varpi}_{k,m,k}-\sum_{k=1}^{K}\sum_{j=1}^{K}b_{k,m,j}^*\boldsymbol{\varpi}_{k,m,j}\label{eq46b},}\\\nonumber
&P=\sum_{k=1}^{K}\sum_{m=1}^{M}\sum_{j=1}^{K}|b_{k,m,j}|^2+\sum_{k=1}^{K}\sum_{m=1}^{M}\boldsymbol{\rho}_{k,m}^H\boldsymbol{\Xi}_{k,m}\boldsymbol{\rho}_{k,m}
\\&-2\sum_{k=1}^{K}\sum_{m=1}^{M}\sqrt{\zeta_{k,m}}\Re\{b_{k,m,k}\}.\label{eq46c}
%&\boldsymbol{\Phi}=[\boldsymbol{\phi}_{1}^{T},\boldsymbol{\phi}_{2}^{T},\cdots,\boldsymbol{\phi}_{M}^{T}]^{T}.
\end{align}
\end{subequations}

{Define $\boldsymbol{\phi}=[\boldsymbol{\phi}_{1}^{T},\boldsymbol{\phi}_{2}^{T},\cdots,\boldsymbol{\phi}_{M}^{T}]^{T}$,
$\boldsymbol{\upsilon}=[\boldsymbol{\upsilon}_{1}^{T},\boldsymbol{\upsilon}_{2}^{T},\cdots,\boldsymbol{\upsilon}_{M}^{T}]^{T}$,
$\boldsymbol{Q}=diag(\boldsymbol{Q}_{1},\boldsymbol{Q}_{2},\cdots,\boldsymbol{Q}_{M})$,}

{Then $f_5$ in (45) can be rewritten in the compact manner.}

{\begin{equation}\label{eq47}
	f_5(\boldsymbol{\phi})=-\boldsymbol{\phi}^H\boldsymbol{Q}\boldsymbol{\phi}
+2\Re\{\boldsymbol{\phi}^H\boldsymbol{\upsilon}\}-P,
\end{equation}}

{Considering (4) and (9d),  $\mathcal{P}(\mathrm{5})$  is equivalent  to  $\mathcal{P}(\mathrm{6})$:}

{\begin{subequations}\label{eq48}
\begin{align}
\nonumber\mathcal{P}(\mathrm{6}):\underset{\boldsymbol{\phi},\varphi_{i,r},\psi_{i,r},\kappa_{i,r}}\min & f_6(\boldsymbol{\phi})=\boldsymbol{\phi}^H\boldsymbol{Q}\boldsymbol{\phi}
-2\Re\{\boldsymbol{\phi}^H\boldsymbol{\upsilon}\}\\
\text{s.t.}~~ &\mathbf{\phi}_{i,r,m}=\frac{\varphi_{i,r}f_{m}^{2}}{\psi_{i,r}^2-f_{m}^{2}+j\kappa_{i,r}f_{m}},\label{eq48a}\\
&|\mathbf{\phi}_{i,r,m}|\leq1,\label{eq48b}\\
&\nonumber\forall i\in\mathcal{N_C},\forall r\in\mathcal{R}, \forall m\in\mathcal{M},
\end{align}
\end{subequations}}

{Due to the complex constraints in (48), problem $\mathcal{P}(\mathrm{6})$ is a difficult optimization problem, and its solution is a challenging task. To solve this issue, we employ the penalty-based approach, whereby the equality constraint (48a) is incorporated into the objective function as a penalty term. This results in the following optimization problem.}
{
\begin{equation}%\label{eq49}
\begin{aligned}
\nonumber\mathcal{P}(\mathrm{7}):\underset{\boldsymbol{\phi},\boldsymbol{\varphi},\boldsymbol{\psi},\boldsymbol{\kappa}}\min & f_7(\boldsymbol{\phi})=f_6(\boldsymbol{\phi})+\frac{1}{2\mu}||\boldsymbol{\phi}-
\mathbf{b}(\boldsymbol{\varphi},\boldsymbol{\psi},\boldsymbol{\kappa})||^2,\\
\text{s.t.}~~&(48b),
\nonumber\forall i\in\mathcal{N_C},\forall r\in\mathcal{R}, \forall m\in\mathcal{M}.
\end{aligned}
\end{equation}
}
{
The vector $\mathbf{b}\in \mathbb{C}^{MRN_{c}\times 1}$ corresponds to the equality constraint of (48a).
$\boldsymbol{\varphi}=[\varphi_{1,1},\cdots,\varphi_{1,R},\cdots,\varphi_{N_c,1},\cdots,\varphi_{N_c,R}]$,
$\boldsymbol{\psi}=[\psi_{1,1},\cdots,\psi_{1,R},\cdots,\psi_{N_c,1},\cdots,\psi_{N_c,R}]$,
and $\boldsymbol{\kappa}=[\kappa_{1,1},\cdots,\kappa_{1,R},\cdots,\kappa_{N_c,1},\cdots,\kappa_{N_c,R}]$.
The parameter $\mu$ is employed to impose a penalty on the violation of the equality constraint (48a).}

{To address $\mathcal{P}(\mathrm{7})$, we propose an efficient algorithm based on APG and Flecher-Reeves conjugate gradient method (FRCG) for solving the passive precoding problem, which will be discussed in the subsequent subsection.}
%=======================================================section4=====================================================================
%\section{ALTERNATING DIRECTION MULTIPLIER METHOD}\label{sec:4}

{\textit{D3}.~\textit{Fix $\boldsymbol{\varphi},\boldsymbol{\psi},\boldsymbol{\kappa}$ and solve $\boldsymbol{\phi}$}}
\

The APG method is employed to effectively tackle the optimization problem of $\boldsymbol{\phi}$ through the iterative process outlined below.
\begin{equation}\label{eq49}
\boldsymbol{\phi}^{j+1}=\boldsymbol{y}^{j}-\frac{1}{\varpi^{j}} \nabla_{\boldsymbol{\phi}} f_{\mathrm{7}}\left(\boldsymbol{y}^{j}\right),% k=1,2,3,\cdots,
\end{equation}
where $\boldsymbol{y}^{j}$ is the extrapolated point.
\begin{equation}\label{eq50}
\boldsymbol{y}^{j}=\boldsymbol{\phi}^{j}+t_{j}\left(\boldsymbol{\phi}^{j}-\boldsymbol{\phi}^{j-1}\right) ,
\end{equation}
with
\begin{equation}\label{eq51}
t_{j}=\frac{d_{j}-1}{d_{j}}, d_{j}=\frac{1+\sqrt{1+4 d_{j-1}^2}}{2}, d_{0}=0,
\end{equation}
and $\varpi^{j}$ is the step size\footnote{According to the Armijo rule, the step size utilized in~\cite{IEEEconf:58} can be determined. However, this approach introduces an additional loop and increases the complexity of the algorithm. To address these limitations, we treat $\varpi^j$ as a constant throughout  the  iterations.}. Let $\boldsymbol{\varsigma}^{j}=\boldsymbol{y}^{j}-\frac{1}{\varpi^{j}} \nabla_{\boldsymbol{\phi}} f_{\mathrm{7}}\left(\boldsymbol{y}^{j}\right)$,  the projection can be obtained using the following formula.
{
\begin{equation}\label{eq52}
	\phi_{i,r,m}^{j+1} = \left\lbrace
	\begin{array}{l}
		{\frac{\varsigma_{i,r,m}^{j}}{|\varsigma_{i,r,m}^{j}|}},~~~~~~~~~~~~~~~|\varsigma_{i,r,m}^{j}|>1, \\
         \varsigma_{i,r,m}^{j},~~~~~~~~~~~~~~~~|\varsigma_{i,r,m}^{j}|\leq1,\\
		e^{j\theta},~~\theta\in[0,2\pi],~~~~\varsigma_{i,r,m}^{j}=0.
	\end{array}
    \right.
\end{equation}
}
{\textit{D4}.~\textit{Fix $\boldsymbol{\phi}$ and solve $\boldsymbol{\varphi},\boldsymbol{\psi},\boldsymbol{\kappa}$}}
\

{After getting $\boldsymbol{\phi}$, solving $\boldsymbol{\varphi},\boldsymbol{\psi},\boldsymbol{\kappa}$ is equivalent to the following optimization.}
{
\begin{equation}%\label{eq53}
\begin{aligned}
\nonumber\mathcal{P}(\mathrm{8}):\underset{\boldsymbol{\varphi},\boldsymbol{\psi},\boldsymbol{\kappa}}\min  f_8(\boldsymbol{\varphi},\boldsymbol{\psi},\boldsymbol{\kappa})=||\boldsymbol{\phi}-
\mathbf{b}(\boldsymbol{\varphi},\boldsymbol{\psi},\boldsymbol{\kappa})||^2,
%\nonumber\forall i\in\mathcal{N_C},\forall r\in\mathcal{R}, \forall m\in\mathcal{M}.
\end{aligned}
\end{equation}
}
{This is an unconstrained optimization problem, which we solve using the FRCG algorithm. The iterations of this method consist of the following steps.}
{
\begin{equation}\label{eq53}
\left\lbrace
\begin{array}{l}
         \mathbf{z}^{j+1}=\mathbf{z}^{j}+\tau_{j}\mathbf{p}^{j},\\
         \mathbf{p}^{j+1}=-\nabla f_{8}(\mathbf{z}^j)+\lambda_{j}\mathbf{p}^{j},\\
         \lambda_{j}=\frac{||\nabla f_{8}(\mathbf{z}^{j+1})||^2}{||\nabla f_{8}(\mathbf{z}^{j})||^2},\\
         \mathbf{p}^{0}=-\nabla f_{8}(\mathbf{z}^{0}).
         \end{array}
    \right.
\end{equation}}
{where $\mathbf{z}$ represents one of $\boldsymbol{\varphi},\boldsymbol{\psi},\boldsymbol{\kappa}$. $\tau_{j}$ is the step size.}

%{The APG-FRCG algorithm for solving  passive precoding $\boldsymbol{\Phi}$ is summarized in Algorithm 2.}

Remark 2: The PDS method [38] was employed to solve  passive precoding {$\boldsymbol{\Phi}$ in the narrowband systems},  which involves iterative solutions for both $\boldsymbol{\Phi}$ and Lagrangian multipliers $\boldsymbol{\chi}=[\chi_1,\cdots,\chi_{N_cR}]^{T}$\footnote{The symbol in [38] is utilized to represent the Lagrangian multipliers.}. Due to the large number of elements in the IRS, the complexity of the PDS method is significantly high. {It is worth noting that the PDS method can not address the passive precoding $\boldsymbol{\Phi}$ in the windband systems.} In contrast, our proposed {APG-FRCG} algorithm offers efficiency and simplicity as it provides closed-form solutions for {$\boldsymbol{\Phi}$} without requiring updates to Lagrangian multipliers $\boldsymbol{\chi}$. Algorithm 2 summarizes the {APG-FRCG} algorithm to solve passive precoding {$\boldsymbol{\Phi}$}.

 %The $\boldsymbol{\Theta}$ and Lagrangian multipliers $\boldsymbol{\chi}=[\chi_1,\cdots,\chi_{N_cR}]^{T}$\footnote{We use the symbol in [42] for the Lagrangian multipliers.} was  solved iteratively and there are no closed-form solutions. Considering the elements of the IRS is very large, and hence, the complexity of the PDS method  is high. Compared to the PDS method, our presented APG algorithm is efficient and simple because $\boldsymbol{\Theta}$  have simple closed-form solutions and do not need to update the Lagrangian multipliers $\boldsymbol{\chi}$.

\begin{algorithm}
	\caption{APG-FRCG algorithm to solve $\mathcal{P}(\mathrm{4})$}
	\begin{algorithmic}
		%\STATE {1: $\mathbf{Input}$: $\mathbf{\delta}$, {$\mathbf{\xi}$}.}
		\STATE {1: Given $\left(\mathbf{W},\boldsymbol{\eta}\right)$.{~Set initial values for  $\boldsymbol{\phi}, \boldsymbol{\varphi},\boldsymbol{\psi}, \boldsymbol{\kappa}$, and tolerance $\varepsilon>0$}.}
		\STATE {2: $\mathbf{Repeat}$}
		\STATE {3: Update $\boldsymbol{\rho}$ by (\ref{eq41});}
		\STATE {4: Update ${\boldsymbol{\phi}}$ by (\ref{eq49})-(\ref{eq52});}
        \STATE {{5: Update $\boldsymbol{\varphi},\boldsymbol{\psi}, \boldsymbol{\kappa}$ by (\ref{eq53}), respectively;}}
		\STATE {6: $\mathbf{Until}$  the problem $\mathcal{P}(4)$ converges.}
	\end{algorithmic}
\end{algorithm}

{To date, we have addressed the joint precoding design for IRS-assisted cell-free networks.
Based on the aforementioned discussion, we present Algorithm 3 as a summary of the CADMM-APG-FRCG algorithm to solve the  problem of joint precoding design.}
\begin{algorithm}
	\caption{Joint precoding design algorithm based on the CADMM-APG-FRCG for $\mathcal{P}(1)$}
	\begin{algorithmic}
		%\STATE {1: $\mathbf{Input}$: $\mathbf{\delta}$, {$\mathbf{\xi}$}.}
		\STATE {1: $\mathbf{Initial}$~$\mathbf{W},~\mathbf{\Phi}$} to a feasible solution.
		\STATE {2: $\mathbf{Repeat}$}
		\STATE {3: Update $\boldsymbol{\eta}$ by (\ref{eq15});}
		\STATE {4: Solve $\mathbf{W}$ according to Algorithm 1;}
        \STATE {5: Solve ${\mathbf{\boldsymbol{\Phi}}}$ based on Algorithm 2;}
		\STATE {6: $\mathbf{Until}$  the problem $\mathcal{P}(1)$ converges.}
	\end{algorithmic}
\end{algorithm}

\subsection{Complexity analysis}
The proposed {CADMM-APG-FRCG} algorithm for joint precoding design comprises three main components: updating variables $\boldsymbol{\eta}$, solving  active precoding $\mathbf{W}$ and passive precoding $\mathbf{\Phi}$. The solution for $\mathbf{W}$ involves updating $\boldsymbol{\delta}$, $\mathbf{W}$, $\mathbf{V}$ and $\mathbf{q}$ using (\ref{eq17}), (\ref{eq27}), (\ref{eq35}) and (24c), respectively. {Similarly, the solution for $\boldsymbol{\Phi}$ requires updating $\boldsymbol{\rho}$,  $\boldsymbol{\phi}$, $\boldsymbol{\varphi},\boldsymbol{\psi}$, and $\boldsymbol{\kappa}$ using (\ref{eq41}),  (\ref{eq49})-(\ref{eq52}) and (\ref{eq53}), respectively}. We evaluate the computational complexity (CC) of the {CADMM-APG-FRCG} algorithm in terms of complex multiplications (CMs).
Since updating $\mathbf{q}$ and $\mathbf{V}$ do not need CMs, we only need to calculate  CMs when updating~$\boldsymbol{\eta}$,  $\boldsymbol{\delta}$, $\mathbf{W}$,  $\boldsymbol{\rho}$,{ $\boldsymbol{\phi}$, $\boldsymbol{\varphi},\boldsymbol{\psi}$, and $\boldsymbol{\kappa}$}, respectively. The corresponding CM values are provided in Table.~I, where $I_{W}$,  {$I_{\boldsymbol{\phi}}$} and {$I_{1}$} represent the iteration numbers for updating $\mathbf{W}$,  $\mathbf{}{\boldsymbol{\phi}}$ and {FRCG algorithm}, respectively.

%We can see that the proposed CADMM-APG algorithm for the joint precoding design consists of three parts, i.e.,  updating variables $\boldsymbol{\eta}$, solving the active precoding $\mathbf{W}$ and the passive precoding $\mathbf{\Theta}$.  Solving $\mathbf{W}$ requires updating $\boldsymbol{\delta}$, $\mathbf{W}$, $\mathbf{V}$ and $\mathbf{q}$ by (\ref{eq17}), (\ref{eq27}), (\ref{eq35}) and (24c), respectively. Similarly, Solving $\mathbf{\Theta}$ requires updating $\boldsymbol{\rho}$ and $\boldsymbol{\theta}$ by (\ref{eq41}) and (\ref{eq48})-(\ref{eq51}), respectively. We measure the computational complexity (CC) of the proposed CADMM-APG algorithm employing  complex multiplications (CMs). Since updating $\mathbf{q}$ do not need CMs, we only need to calculate the CMs updating~$\boldsymbol{\eta}$,  $\boldsymbol{\delta}$, $\mathbf{W}$, $\mathbf{Z}$, $\boldsymbol{\rho}$ and $\boldsymbol{\theta}$, respectively.
%We provide the CMs in the Table.~I, where $I_{W}$ and $I_{\boldsymbol{\theta}}$ denote the iteration numbers of updating $\mathbf{W}$ and $\mathbf{}\boldsymbol{\theta}$.
\begin{table}[h]
	\begin{center}
		\caption {COMPUTATIONAL COMPLEXITIES}\label{TableI}
		\begin{tabular}{|c|c|}%p{38pt}|}
			
			\hline    Variables                    &Computational complexities $\mathcal{O}\left ( \cdot  \right)$\\
			\hline    $\boldsymbol{\eta}$        &$MK(KN_rN_tN_b+N_r^3+(K+1)N_r^2+N_r)$   \\
            \hline    $\boldsymbol{\delta}$        &$MK(KN_rN_tN_b+N_r^3+(K+1)N_r^2)$     \\
			\hline    $\mathbf{W}$                   &$N_t^{2}N_b^{2}M^{2}K^{2}+I_{W}N_tN_bMK$    \\
		    %\hline    $\mathbf{V}$                   &$I_{W}N_tN_bZK$     \\
            \hline    $\boldsymbol{\rho}$       &$MK(N_r^3+(K+1)N_r^2)$    \\
            \hline    {$\boldsymbol{\phi}$}       &{$I_{\boldsymbol{\phi}}(N_{c}^{2}R^2+2N_cR)$}     \\
			\hline    {$\boldsymbol{\varphi}$   }   &{$3I_{1}N_cR$  }        \\
            \hline    {$\boldsymbol{\psi}$ }     &{$3I_{1}N_cR$ }         \\
            \hline    {$\boldsymbol{\kappa}$}      &{$3I_{1}N_cR$}          \\
            \hline
		\end{tabular}
	\end{center}
\end{table}

Given the substantial number of BSs, users, and IRS elements in the IRS-assisted cell-free network, it is reasonable to assume that $N_tN_bMK\gg N_r$ and $N_cR\gg N_r$ [9], [38], [39]. Based on this assumption and Table.~I, it can be observed that the CCs of the overall algorithm  depend primarily on those associated with the calculation of $\mathbf{W}$ and ${\boldsymbol{\Phi}}$. Consequently, the CC of the overall algorithm can be expressed as {$\mathcal{O}(I_{oC}(N_t^{2}N_b^{2}M^{2}K^{2}+I_{W}N_tN_bMK+I_{\boldsymbol{\phi}}(N_{c}^{2}R^2+2N_cR$
\\
$+9I_{1}N_cR)))$},%\footnote{Here we use the fact that
%$R^{3}N_c(K^2+M)\gg I_{\boldsymbol{\hat{\theta}}}M_{j}R(I_1+I_2)$.},\\%+2I_{\boldsymbol{\theta}}N_cR)\right)$,
~where $I_{oC}$ represents the number of iterations for the {CADMM-APG-FRCG} convergence.%\footnote{For the update of $\mathbf{W}$ in (27), the term $\mathbf{D}\mathbf{W}_0$ only needs to be calculated  once.}. In particular, compared to PDS method, our proposed CADMM-APG algorithm exhibits significantly lower CCs, as demonstrated in Section IV.

%Because the number of BSs,  users and  IRS elements in the IRS-assisted cell-free network is large [9], [42], [43],  $N_tN_bZK\gg N_r$ and $N_cR\gg N_r$ are  reasonable assumptions. Based on the assumption and  Table.~I, we see that  CCs of the overall algorithm  depend mainly on the CCs for calculating $\mathbf{W}$, $\mathbf{V}$, $\mathbf{}\boldsymbol{\theta}$ and $\mathbf{\psi}$. Therefore, the CC of the overall algorithm is  $\mathcal{O}\left(I_{oCL}(N_t^{2}N_b^{2}Z^{2}K^{2}+2I_{W}N_tN_bZK+N_c^{2}R^2+2I_{\boldsymbol{\theta}}N_cR)\right)$, where $I_{oCL}$ is the iteration number for the CADMM-APG convergence\footnote{For the updates of $\mathbf{W}$ in (27), the terms $\mathbf{D}\mathbf{W}_0$ only need to be calculated  once.}.   The CC of the CADMM-APG algorithm is much lower than that of the PDS method, which will be verified in Section IV.

%which appear in () most complex operation in the ADMM algorithm is updating $\mathbf{Z}^{j+1}$  and $\mathbf{r}^{j+1}$ . The matrix inverse terms in both $\mathbf{Z}^{j+1}$  and $\mathbf{r}^{j+1}$  can be precomputed in the initialization step. Therefore, the complexity of ADMM algorithm is $\mathcal{O}(N^3)$ , which is much lower than $\mathcal{O}(N^6)$  of PDS in [40]. In addition, through parallel computation, especially when N is large, ADMM algorithm converges faster.
%=======================================================section5=====================================================================
\section{SIMULATION RESULTS}\label{sec:5}
\subsection{Simulation Configuration}
In order to assess the performance of the {CADMM-APG-FRCG} algorithm in a cell-free network with IRS assistance, we conducted simulations under various conditions. We followed the deployment scheme used in previous studies [4], [38] (see Fig. 5 in [38]), where four users were simultaneously served by five BSs. To address capacity limitations caused by obstructions in green areas, we placed two IRSs independently on different building surfaces. For our simulation setup, we considered a three-dimensional coordinate system. The j-th BS was positioned at coordinates (40(j-1)m, -50m, 3m), while the IRSs were located at coordinates {(30m, 10m, 6m) and (130m, 10m, 6m)}. We incorporated a distance-dependent path loss model to account for propagation loss.

\begin{equation}\label{eq54}
L(\tilde{d})\!=c_{0}\left(\frac{\tilde{d}}{d_{0}}\right)^{-\zeta},
\tilde{d}\!\in \{\tilde{d}_{BU}, \tilde{d}_{IU}, \tilde{d}_{BI}\}, \!\zeta\!\in \{\zeta_{BU}, \zeta_{IU}, \zeta_{BI}\}
\end{equation}
where $c_0=-30$dB represents the path loss at the reference distance $d_0=1$ m, $\zeta$ denotes  the path loss exponent, and $\tilde{d}$ indicates the link distance. The link distances between the BS and user, IRS and user,  BS and IRS  are denoted by $\tilde{d}_{BU}$, $\tilde{d}_{IU}$, and $\tilde{d}_{BI}$, respectively. $\zeta_{BU}, \zeta_{IU}$, and $\zeta_{BI}$ are corresponding path loss exponent of them, respectively. {In the following simulations, we take $\zeta_{BU}=3.5$, $\zeta_{IU}=2.8$, and $\zeta_{BI}=2.2$.} In this cell-free network scenario, each BS is equipped with a limited number of antennas and operates at low transmit power. Therefore, we set the maximum transmit power of each BS as $P_{n_b,max}=0$ dBm. The number of antennas per BS is $N_t=2$, while each user has $N_r=2$ antennas. Additionally, there are $R=100$ units per IRS and {$M=16$} tones in use. {The Lorentzian parameters were initialized as:~$\varphi_{i,r}=1$,~$\psi_{i,r}=3\times10^9$,~$Q_n=50$, ~$\kappa_{i,r}=\frac{\psi_{i,r}}{Q_n}=6\times10^7$,~$f_c=3$ GHz and $B=100$ MHz.} The noise power $\tau^2 $ is assumed to be $-80$ dBm [25]. We adopt the Rice fading channel model for our analysis.

\begin{equation}\label{eq55}
\begin{aligned}
&\mathbf{H}=\sqrt{\frac{\varepsilon}{1+\varepsilon}}\mathbf{H}_{LoS}+\sqrt{\frac{1}{1+\varepsilon}}\mathbf{H}_{NLoS}, \\ ~~&\mathbf{H}\in\{\mathbf{H}_{BU},\mathbf{H}_{IU},\mathbf{H}_{BI}\}; \varepsilon\in\{\varepsilon_{BU}, \varepsilon_{IU}, \varepsilon_{BI}\},
\end{aligned}
\end{equation}
where $\varepsilon$ denotes the Rice factor.  $\mathbf{H}_{LoS}$ and $\mathbf{H}_{NLoS}$ stand for the line-of-sight (LoS)  and Rayleigh fading component, respectively.  {When $\varepsilon\rightarrow\infty$, $\mathbf{H}$ corresponds to a LoS channel. As $\varepsilon\rightarrow0$, $\mathbf{H}$ is a Rayleigh fading channel.  In addition, we  assume $\varepsilon_{BU}\rightarrow0$, $\varepsilon_{BI}\rightarrow\infty$ and $\varepsilon_{IU}\rightarrow0$ in simulations. The weight factor is assumed to be $\xi_{k,m}=1$. In the CADMM-APG-FRCG algorithm, we  select the parameters $\alpha=N_b$, $\beta=\frac{\mathbf{W}^{H}\mathbf{D}\mathbf{W}}{\mathbf{W}^{H}\mathbf{W}}$, $\varpi=\frac{1}{8}$, and $\mu=12\frac{N_{b}^{2}}{\boldsymbol{\phi}^H\mathbf{Q}\boldsymbol{\phi}}$.}

We also assume that the users are uniformly distributed within a circular area with a radius of ${10}$ m, centered at $(L, 0)$, and that all users have a height of $1.5$ m. To evaluate the performance of the proposed CADMM-APG-FRCG algorithm compared to the PDS method [38] in an IRS-assisted cell-free system, we consider four baselines.

%$\textbf{Ideal IRS case}$: Optimizing the WSR based on the perfect CSI scenario.

$\textbf{Optimized phase shift}$:  Maximizing the WSR is achieved through Algorithm. 3.

$\textbf{Without direct link}$: Maximizing the WSR is achieved via Algorithm. 3, where direct links between BSs and users are obstructed, leaving only indirect channels available.

$\textbf{Random phase shift}$: Maximizing the WSR only at BSs based on  Algorithm. 3 by randomly setting the phase shift of all IRS reflection units.

$\textbf{Without IRS}$:  By utilizing the CSI between BSs and users, the optimization of WSR is executed through a precoding method implemented at the BSs.
\subsection{Performance of the Proposed  CADMM-APG-FRCG Algorithm }

{Fig.~2 illustrates that both the PDS and CADMM-APG-FRCG algorithms exhibit two peaks of the WSR at $L=30$ and $L=130$, except for scenarios without IRS or with random phase shift. This observation suggests that the WSR improves as users approach IRSs. Nevertheless, when a random phase shift is considered for the wideband case, there is only a modest increase in WSR in comparison to scenarios without IRS. In contrast, the narrowband case exhibits minimal variation.  These findings indicate that proper optimization of passive beamforming can significantly enhance WSR in cell-free networks through the use of IRS. Moreover, it is noteworthy that, with the exception of the cases without IRS and with random phase shift, across all  scenarios considered, the CADMM-APG-FRCG algorithm consistently outperforms the PDS algorithm,  demonstrating a more substantial improvement in WSR performance.} This validates the effectiveness of the CADMM-APG-FRCG algorithm in maximizing WSR in the presence of IRS.
\begin{figure}
	\centering
	\includegraphics[width=3.5in,angle=0]{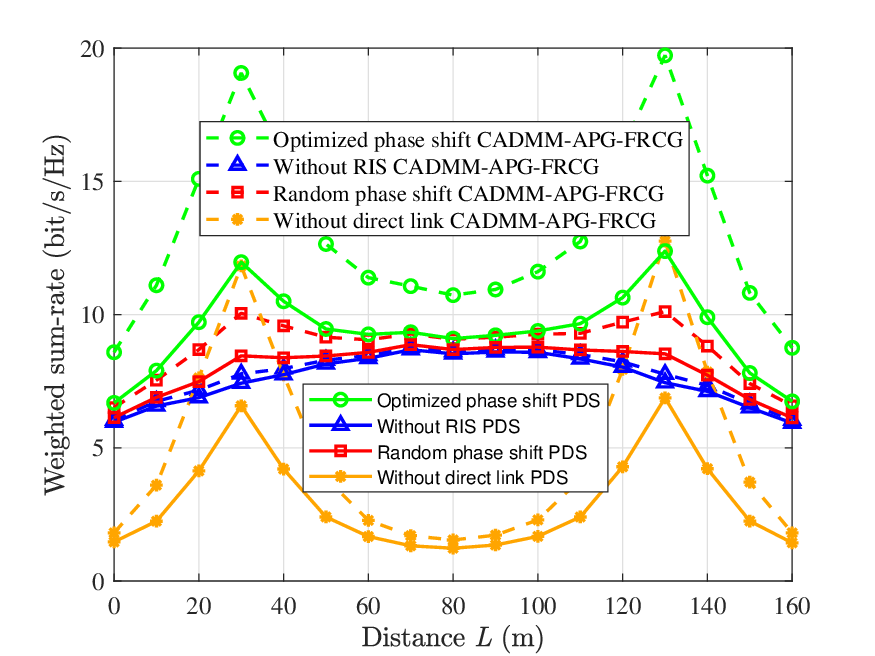}
	\caption{Weighted sum-rate against the distance $L$.}
	\label{fig2}
\end{figure}

Fig.~3 demonstrates the convergence of the {CADMM-APG-FRCG} algorithm and the PDS method in a cell-free system with IRS assistance. The horizontal axis represents the number of iterations. When setting {$L=30$} and keeping other parameters constant, it is evident that across three scenarios, namely the ideal case, the continuous phase shift case and the case without a direct link, the {CADMM-APG-FRCG} algorithm consistently outperforms the PDS method in terms of WSR values. From Fig.~3, we can see that the PDS algorithm requires $15$ iterations to converge, whereas the {CADMM-APG-FRCG} algorithm requires only {$10$} iterations to achieve convergence. {As the {CADMM-APG-FRCG} algorithm necessitates a lesser number of iterations than the PDS algorithm, its complexity is considerably lower than that of the PDS algorithm, as illustrated in Table.~II below.  This markedly reduces the time required for simulations to address problem  $\mathcal{P}(\mathrm{0})$, thereby rendering our {CADMM-APG-FRCG} algorithm a viable option for engineering applications. Furthermore, it is evident that upon convergence, the WSR obtained by the {CADMM-APG-FRCG} algorithm is approximately $62.8\%$ greater than that of the PDS algorithm. This indicates that the PDS algorithm only attains a suboptimal solution. The underlying reason for this is that the {CADMM-APG-FRCG} algorithm optimally solves $\mathbf{W}$ and $\boldsymbol{\Phi}$ in $\mathcal{P}(\mathrm{0})$}.%However, because the complexity of the LADMM algorithm is much lower than that of the PDS algorithm, the LADMM algorithm has faster convergence speed.
\begin{figure}
	\centering
	\includegraphics[width=3.5in,angle=0]{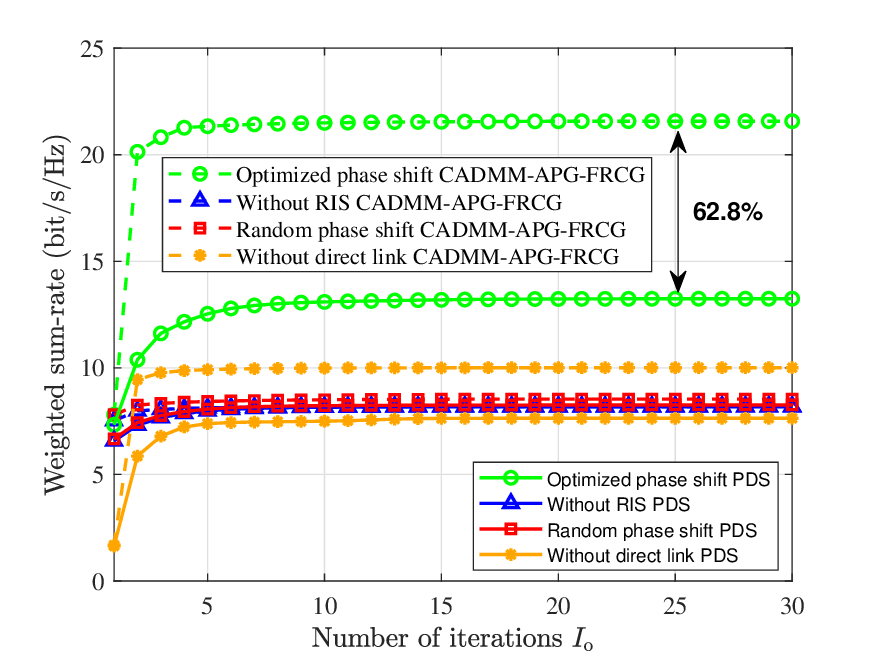}
	\caption{Weighted sum-rate against the number of iterations $I_o$.}
	\label{fig3}
\end{figure}

It is challenging to obtain accurate CSI in IRS-assisted cell-free networks, even with the assumption of perfect CSI. In Fig.~4, we investigate the relationship between the WSR and the error in channel estimation.  We model the actual estimated channel as follows [38].

\begin{equation}\label{eq56}
\tilde{\mathbf{H}}=\mathbf{H}+\mathbf{\triangle H},
\end{equation}

The practical channel is represented by $\mathbf{H}$, and the estimation error is denoted as $\mathbf{\triangle H}$, where $\mathbf{\triangle H}\sim \mathcal{CN}(0, \tau_{H}^{2}\mathbf{I})$. To characterize the CSI estimation error, we assume $\tau_{H}^{2}=\omega||\mathbf{H}||^2$. As illustrated in Fig. 4,  the  degradation of WSR performance is evident as the  error  $\omega$ increases. In comparison to the scenario of perfect CSI $(\omega=0)$, when  $\omega=0.2$, the PDS algorithm exhibits a WSR performance loss of approximately $15\%$, while CADMM-APG-FRCG shows a loss of about $14\%$. With $\omega=0.3$, the WSR performance loss for the PDS algorithm and CADMM-APG-FRCG reaches about $25\%$ and $25\%$, respectively. It can be observed that  both algorithms exhibit similar resilience to CSI estimation errors. However, it is worth noting that the WSR obtained by the CADMM-APG-FRCG algorithm is $66.6\%$ and $65.7\%$ higher than that of the PDS algorithm when $\omega=0.2$ and $\omega=0.3$, respectively.

\begin{figure}
	\centering
	\includegraphics[width=3.5in,angle=0]{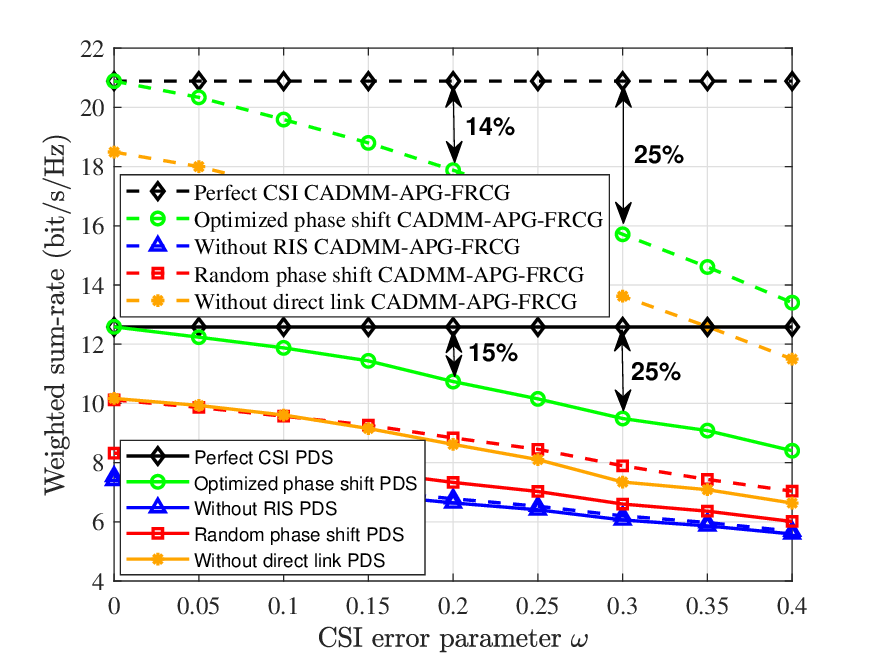}
	\caption{Weighted sum-rate against the CSI error parameter $\omega$.}
	\label{fig4}
\end{figure}

\begin{figure}
	\centering
	\includegraphics[width=3.5in,angle=0]{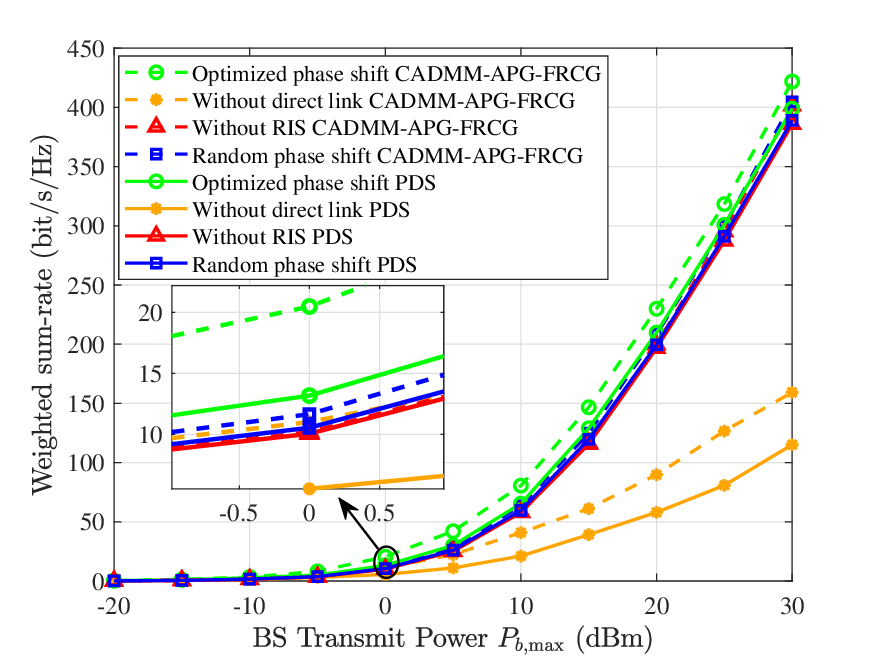}
	\caption{Weighted sum-rate against the BS transmit power $P_{\mathrm{n_b,max}}$.}
	\label{fig5}
\end{figure}

To examine the influence of various system parameters on system performance, we relocated four users to positions {$(20m, 0)$, $(60m, 0)$, $(100m, 0)$, $(140m, 0)$}. Fig.~5 depicts the relationship between the average WSR and the BS transmission power. {~It can be observed that an increase in the BS transmit power results in an elevated average WSR. Intuitively, the similarity between the CADMM-APG-FRCG and PDS algorithms lies in the fact that when the BS transmission power is insufficient, the signal reflected by the RIS is very weak and does not significantly contribute to performance improvement. Conversely, when the BS transmission power is excessive, the BS tends to allocate most of the power to the beam aimed at the BS-user link rather than the BS-RIS-user link, making the role of RISs less significant. This explains the difference between the "Without RIS" and "Without direct link" scenarios, and as the BS transmission power increases, the gap between the two methods becomes more pronounced. Therefore,  a better selection of BS transmission power is crucial to fully leverage the use of RIS in RIS-aided cell-free networks. However, it is noteworthy that among all the aforementioned schemes, the PDS method yields suboptimal solutions, whereas the CADMM-APG-FRCG method achieves optimal solutions.  Consequently, the CADMM-APG-FRCG algorithm outperforms the PDS algorithm in terms of WSR performance, when the BS transmission power is large or equal to 0 dBm.}
\begin{table*}[tbp]
	\centering
    \setlength\tabcolsep{3pt}
	\caption {Computational complexity comparison between the CADMM-APG and PDS algorithms}\label{TableII}
	\begin{tabular}{|c|c|c|c|}
		\hline
		Methods     & CC $\mathcal{O}\left ( \cdot  \right)$ & Number of iterations  &  Overall CC \\ \hline
		{CADMM-APG-FRCG}     &       $\begin{aligned}
			& \mathcal{O}\left(I_{oC}(N_t^{2}N_b^{2}M^{2}K^{2}+I_{W}N_tN_bMK+ \right.\\
			& \left.{I_{\boldsymbol{\phi}}(N_{c}^{2}R^2+2N_cR+9I_{1}N_cR)})\right)
		\end{aligned}$
		 & { $I_{oC}=10$, $I_{W}=35$, $I_{\boldsymbol{\phi}}=40$, $I_{1}=5$} &      {2.408E+7 }              \\ \hline
		PDS    &        $\mathcal{O}\left(I_{o}(I_{a}N_t^{2}N_b^{2}M^{2}K^{2}+I_{p}N_c^{2}R^2)\right)$  &  $I_{o}=15$, $I_{a}=11$, $I_{p}=15$  &     7.6584E+7             \\ \hline
		%AO     &    $\mathcal{O}\left(I_\mathbf{o}\left(I_\mathbf{\mu1}I_\mathbf{W1}KN^3+I_\mathbf{R}K^2M^2\right)\right)$  &  $I_\mathbf{\mu1}=20$, $I_\mathbf{W1}=20$,$I_\mathbf{R}=2$  & 4.224E+5                       \\ \hline
		%ICU       &        $\mathcal{O}\left(I_\mathbf{o}\left(I_\mathbf{\mu2}I_\mathbf{W2}KN^3+K^2M^2\right)\right)$ &  $I_\mathbf{\mu2}=22$, $I_\mathbf{W2}=22$   &     2.839E+5          \\ \hline
		%ADMM       &        $\mathcal{O}\left(I_\mathbf{o}\left(I_\mathbf{\mu3}I_\mathbf{W3}KN^3+I_\mathbf{t}M^2+M^3\right)\right)$ &  $I_\mathbf{\mu3}=21$, $I_\mathbf{W3}=21$, $I_\mathbf{t}=100$   &     2.113E+6          \\ \hline
\multicolumn{3}{|c|}{$\text{CCR}$}                & {31.4426\%  } \\ \hline
%\multicolumn{3}{|c|}{$\frac{\mathcal{O}(MD\ and\ APG)}{\mathcal{O}(AO)}$}                & 41.1018\%   \\ \hline
%\multicolumn{3}{|c|}{$\frac{\mathcal{O}(MD\ and\ APG)}{\mathcal{O}(ICU)}$}                & 61.0285\%   \\ \hline
	\end{tabular}
\end{table*}

%When the transmission power is $30$ dB, there is almost no difference in the WSR between the two methods in all cases, and the WSR of the CADMM-APG is only slightly better than that of the  PDS. The reason for this phenomenon is that when the BS transmission power is too low, the signal reflected by the  IRS is relatively weak, so  the impact on the system can be ignored. At this point, it is equivalent to only optimizing the traditional cell-free network, i.e., solving the problem $\mathcal{P}(\mathrm{2})$. Due to the fact that the PDS method only obtains a suboptimal solution, while the CADMM-APG method obtains an optimal solution, the CADMM-APG method has better WSR performance than the PDS method. When the BS transmission power  is very high, the BS tends to allocate more power to the BS-User direct link instead of the BS-IRS-User link, resulting in a less significant IRS effect.

The relationship between the number of IRS units and the WSR is illustrated in Fig. 6, using the same settings as Fig.~5. {As can be observed in Fig. 6, the average WSR of IRS-assisted cell-free systems increases with an increase in the number of IRS units in all three cases, namely, the ideal case, continuous phase shift case and without direct link case. In cases involving random phase shift and without IRS, the incorporation of additional IRS units exerts a negligible influence on the WSR performance, as observed in both the PDS and CADMM-APG-FRCG methods.  However, for the two methods, random phase shift results in a slightly superior WSR in comparison to the case without IRS. It should be noted that since $\boldsymbol{\phi}$ has a dimension of $MRN_c$, the calculation of the WSR becomes more complex with an increasing number of IRS units. This, in turn, presents a challenge in channel estimation.  Therefore, it is necessary to strike a balance between achieving optimal WSR performance and determining an appropriate number of IRS units to use. Furthermore, it has been observed that the CADMM-APG-FRCG algorithm significantly outperforms the PDS algorithm in terms of the WSR obtained.}
\begin{figure}
	\centering
	\includegraphics[width=3.5in,angle=0]{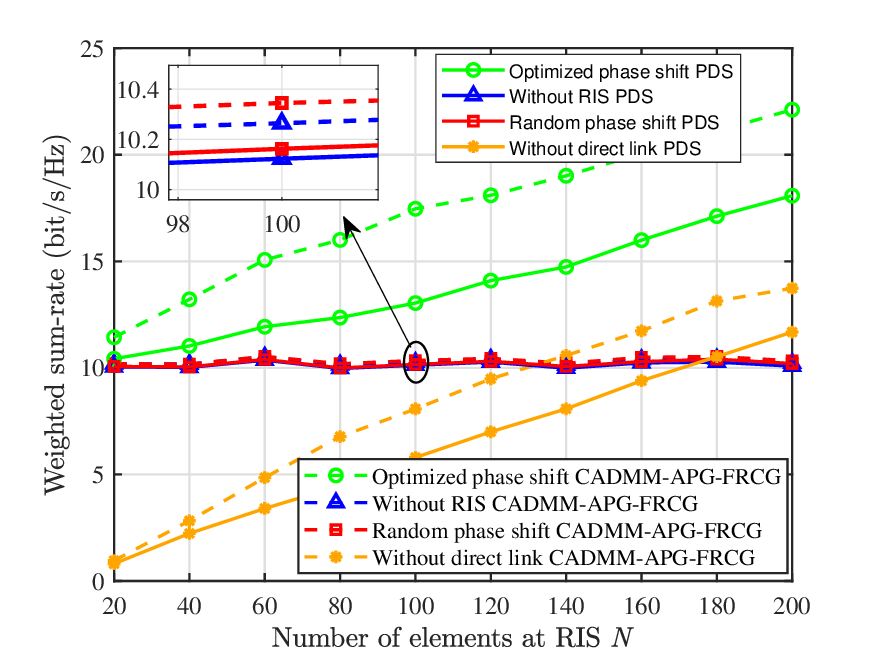}
	\caption{Weighted sum-rate against the number of RIS elements.}
	\label{fig6}
\end{figure}
\begin{figure}
	\centering
	\includegraphics[width=3.5in,angle=0]{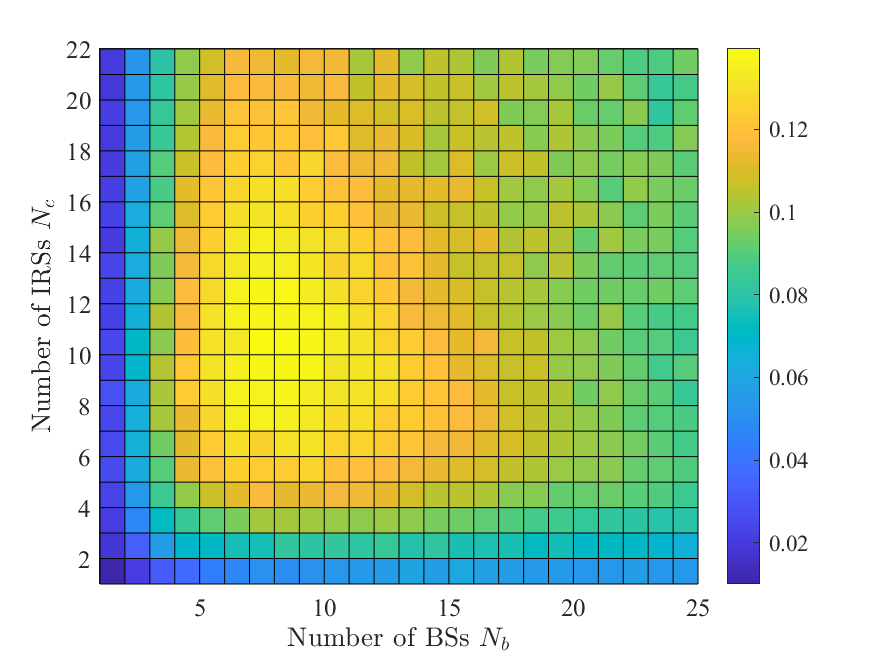}
	\caption{Energy efficiency of the system in CADMM-APG-FRCG method against BS number $N_b$ and IRS number $N_c$.}
	\label{fig7}
\end{figure}

The computational complexity  between the {CADMM-APG-FRCG} and PDS algorithms is compared in Table II\footnote{For the complexity of the PDS method, we use the symbol in [38]. Here, $I_{oCA}$, $I_{W}$, $I_{\boldsymbol{\phi}}$, $I_1$, $I_{o}$, $I_{a}$ and $I_{p}$ are based on the average of 100 channel experiments.}. To illustrate the low CC property of the {CADMM-APG-FRCG},  we provide the computational complexity ratio (CCR) of the {CADMM-APG-FRCG} algorithm versus the PDS one. The CCR is defined as follows.
\begin{equation}\label{eq58}
\textrm{CCR}=\frac{\text{the overall CC of {CADMM-APG-FRCG}}}{\text{the overall CC of PDS}}.
\end{equation}

{The  $\textrm{CCR}=\frac{2.408E+7}{7.6584E+7}= 31.4426\%$, which indicates a $68.5574\%$} lower CC compared to that of the PDS. Based on Fig.~2-Fig.~6 and Table. II, it can be observed that the {CADMM-APG-FRCG} algorithm exhibits superior WSR performance and lower computational complexity when solving the problem $\mathcal{P}(\mathrm{1})$ compared to the PDS algorithm. Therefore, the proposed {CADMM-APG-FRCG} method presents itself as a highly promising approach for addressing complex optimization problems  $\mathcal{P}(\mathrm{1})$.

In Fig.~7, we investigate the correlation between the energy efficiency of the {CADMM-APG-FRCG} algorithm and the number of BS and IRS, with the aim of improving the capacity of the   system through  the use of IRS with low power consumption and  cost-effectiveness.  The energy efficiency of the system is quantified using a formula derived from references [22] and [38].%aiming to enhance system capacity through cost-effective and low-power consumption IRS utilization.
\begin{equation}\label{eq58}
E_{ee}=\frac{R_{sum}}{\hat{\lambda}||\mathbf{W}||+N_bP_{B}+KP_{U}+N_cRP_{I}},
\end{equation}
where $\hat{\lambda}^{-1}$ denotes the efficiency of the transmit power amplifier, and $P_{B}$, $P_{U}$,  and $P_{I}$  represents the power consumption at each BS, each user and each IRS unit, respectively. We adopt the same configuration as the reference [38], that is, $\lambda=1.2$ , $P_{B}=9$ dBW, $P_{U}=10$ dBm, $P_{I}=10$ dBm, $M=N_t=N_r=1$  and $R=20$, with the corresponding distances set to $d_{BU}=d_{BI}=110$ m and $d_{IU}=15$ m.  As illustrated in Fig. 7,  the  energy efficiency does not increase monotonically  with the growth of $N_b$ and $N_c$.  When $N_b$ is held constant, an initial enhancement in energy efficiency is followed by a subsequent decrease as $N_c$ increases.  {When $N_b=7$  and $N_c=10$, $E_{ee}$ reaches its maximum value at this point with a value of $0.1394$ bit/s/Hz/W being achieved.  However, when $N_b=7$ and $N_c=22$, $E_{ee}$ drops to $0.1129$ bits/s/Hz/W, resulting in a decrease of approximately $19.01\%$ in energy efficiency. In comparison to the PDS method\footnote{The maximum energy efficiency of the PDS method is $0.138$ bit/s/Hz/W in [38], the corresponding decrease in energy efficiency is $34.7\%$.}, the CADMM-APG-FRCG algorithm demonstrates a more gradual decline in energy efficiency than the PDS algorithm. This offers a broader range of possibilities for the selection of energy efficiency and the number of BSs and IRSs in a practical system.}

%---------------------------------------------------------------------------------------------------------------------------------------------section5
\section{CONCLUSION}

This paper investigates the use of IRS-assisted cell-free networks in a wideband scenario to enhance network capacity while minimizing costs and power consumption. The aim is to maximize the system's WSR by jointly optimizing the precoding design, considering power constraints at BSs and phase shift constraints at the IRSs. To solve this non-convex optimization problem, the paper employs the Lagrangian dual transform to decouple the problem and utilizes fractional programming to address the active and passive beamforming optimizations. Specifically, the {CADMM-APG-FRCG} algorithm is proposed to solve the corresponding subproblems. Simulation results demonstrate that the {CADMM-APG-FRCG} scheme outperforms the primal-dual subgradient (PDS) method in terms of WSR performance while maintaining a lower complexity. This indicates that the {CADMM-APG-FRCG} algorithm offers significant advantages for optimizing network capacity in IRS-assisted cell-free networks.

%\begin{figure} \centering
%\includegraphics[width=6in,angle=0]{differentiteration.eps}
%\caption{Comparison of PAPR reduction for different methods, W=2.}
%\label{fig5}
%\end{figure}

%\begin{figure} \centering
%\includegraphics[width=6in,angle=0]{plotallpaprw4sub8.eps}
%\caption{Comparison of PAPR reduction for different methods, W=4.}
%\label{fig6}
%\end{figure}

%\begin{figure} \centering
%\includegraphics[width=6in,angle=0]{ABCPSOGA.eps}
%\caption{Comparison of PAPR reduction for different methods, W=2.}
%\label{fig7}
%\end{figure}

\end{document}